\documentclass[range]{ar2e}

\usepackage{amssymb}
\usepackage{amsmath}
\usepackage[dvips]{graphicx,epsfig,color}
\usepackage{axodraw4j}
\usepackage{subfigure}
\usepackage{cite}

\newcommand{\mass}{{\mathcal{M}}}
\newcommand{\Imag}{{\rm Im}}
\newcommand{\Real}{{\rm Re}}
\newcommand{\base}{\baselineskip 16pt}

\begin{document}

\input epsf.tex    

\input psfig.sty

\title{
\vspace{-2.2cm} \hspace{10cm}{\small \hfill{CPT/10/18}}\\[-1.0cm]\hspace{10cm}{\small \hfill{DESY 10-016}}\\[-1.0cm]\hspace{10cm}{\small \hfill{IPPP/10/09}}\\[0.2cm]
The Low-Energy Frontier of Particle Physics}

\markboth{The Low-Energy Frontier of Particle Physics}{The Low-Energy Frontier of Particle Physics}

\author{Joerg Jaeckel
\affiliation{ Institute for Particle Physics Phenomenology, Durham University, Durham DH1 3LE, United Kingdom}
Andreas Ringwald
\affiliation{Deutsches Elektronen-Synchrotron, Notkestra\ss e 85, D-22607 Hamburg, Germany}}


\begin{abstract}
\base Most embeddings of the Standard Model into a more unified
theory, in particular the ones based on supergravity or
superstrings, predict the existence of a hidden sector of particles
which have only very weak interactions with the visible sector
Standard Model particles. Some of these exotic particle candidates
(such as e.g. ``axions", ``axion-like particles" and ``hidden U(1)
gauge bosons") may be very light, with masses in the sub-eV
range, and have very weak interactions with photons.
Correspondingly, these very weakly interacting sub-eV particles
(WISPs) may lead to observable effects in experiments (as well as in
astrophysical and cosmological observations) searching for light
shining through a wall, for changes in laser polarisation, for
non-linear processes in large electromagnetic fields and for
deviations from Coulomb's law. We present the physics case and a
status report of this emerging low-energy frontier of fundamental
physics.\\[5ex]
\end{abstract}

\maketitle

\base
\section{Introduction}

We are entering exciting times in particle physics: the Large Hadron Collider
(LHC) is setting a new landmark at the high-energy frontier and probes,
through the collision of multi-TeV protons, the structure of
matter and space-time at an unprecedented level. There is a lot of circumstantial
evidence that the physics at the TeV scale exploited at LHC will bring decisive insights into fundamental
questions such as the origin of particle masses, the nature of dark matter in the universe, and
the unification of all forces, including gravity.
Indeed, most proposals to embed the Standard Model of particle physics into a more general,
unified framework, notably the ones based on string theory or its low energy incarnations,
supergravity and supersymmetry, predict new heavy, $m\gtrsim 100$~GeV,
particles which may be searched for at TeV colliders. Some of these particles,
prominent examples being neutralinos, are natural candidates for the constituents of
cold dark matter in the form of so-called weakly interacting massive particles (WIMPs).

However, there is also evidence that there is fundamental physics at the sub-eV scale.
Indeed, atmospheric, reactor, and solar neutrino data strongly support the hypothesis
that neutrinos have masses in the sub-eV range. Moreover, the vacuum energy density of
the universe, as inferred from cosmological observations, points to the sub-eV range,
$\rho_\Lambda \sim {\rm meV}^4$. As a matter of fact, many of the above mentioned extensions
of the Standard Model not only predict WIMPs, but also WISPs -- very weakly interacting
sub-eV particles. Prominent candidates for such particles go under the
names axions and axion-like particles, often arising as Nambu-Goldstone bosons associated with
the breakdown of global symmetries. Further WISP candidates are massless or light extra, hidden U(1)
gauge bosons as well as light, chiral fermions charged under this hidden U(1). These particles
are frequently encountered in string embeddings of the Standard Model.
The latter potentially contain also light moduli fields and light gravitinos as further
WISP candidates.

Unlike for WIMPs, TeV colliders are not the best means to search for WISPs.
For this purpose, low energy experiments exploiting lasers, microwave cavities, strong electromagnetic
fields, torsion balances etc. seem to be superior.
It is the purpose of this review to present the physics case and a status report
of this emerging low-energy frontier of fundamental physics.

The organization of this review is as follows. In the following Sect.~\ref{physicscase}
we will argue that
many extensions of the Standard Model predict new particles and phenomena at low energies.
On the one hand new light particles are suggested to solve puzzling experimental
results but on the other hand they also appear to be a generic feature of underlying fundamental theories such as string theory.
In Sect.~\ref{astro} we will turn to current constraints from astrophysics and cosmology. Moreover, we will discuss a
few interesting observations that could be explained by invoking WISPs.
Then, in Sect.~\ref{searches}, we will explore how WISPs can be searched for in a variety of controlled laboratory
experiments. We discuss the advantages of these experiments as well as the challenges they face.
Finally in Sect.~\ref{conclusions} we will summarize the current situation and give an outlook towards the future.

\section{Physics Case for WISPs}\label{physicscase}

\subsection{Axions and Axion-Like Particles}

\subsubsection{The Strong CP Problem and Axions\label{Sec:QCD_axion}}

Quantum chromodynamics (QCD), the non-Abelian gauge theory describing strong
interactions, allows for a CP-violating term in the Lagrangian,
\begin{eqnarray}
{\mathcal L}_{\rm CP-viol.} =
\frac{\alpha_s}{4\pi}\, \theta\, {\rm tr}\, G_{\mu\nu} {\tilde G}^{\mu\nu} \equiv
\frac{\alpha_s}{4\pi}\, \theta\,
\frac{1}{2}\,\epsilon^{\mu\nu\alpha\beta}\,{\rm tr}\, G_{\mu\nu} G_{\alpha\beta},
\label{topterm}
\end{eqnarray}
where $G$ is the gluonic field strength. Similar to the strong coupling constant $\alpha_s$,
the fundamental parameter $\theta$ has to be determined experimentally.
One of the most sensitive probes for it is the electric dipole moment of the neutron, arising from the CP-violating term
given in Eq.~(\ref{topterm}). It should be of order
\begin{eqnarray}
\left| d_n\right| \sim \frac{e}{m_n} \left( \frac{m_q}{m_n}\right)
\left|\bar{\theta}\right|
\sim 10^{-16}\ \left|\bar{\theta}\right| \ e\,{\rm cm},
\end{eqnarray}
where $m_n$ ($m_q$) is the neutron (a light-quark mass), $e$ is the unit electric charge,
and
\begin{equation}
\bar{\theta} \equiv \theta + {\rm arg\ det\ }M,
\end{equation}
with $M$ being the quark mass matrix. $\bar{\theta}$ is the actual physical CP-violating parameter in the Standard Model.
The current experimental
upper bound on $\left|d_n\right|<2.9\times 10^{-26}\ e\,$cm~\cite{Amsler:2008zzb} places an extremely stringent limit on
\begin{equation}
\left|\bar\theta\right| \lesssim 10^{-10}.
\end{equation}
The strong CP problem is the lack of an explanation why the dimensionless parameter $\bar\theta$,
a sum of two contribution of very different origins, is so unnaturally small.

The axion occurs in course of a possible solution of this problem. In essence, the proposal
of Peccei and Quinn~\cite{Peccei:1977hh} was to promote $\theta$ to a dynamical field which can relax
spontaneously to zero. The axion field $a$ is introduced as a dynamical $\theta$ parameter,
which has a shift symmetry,
\begin{eqnarray}
a\to a + {\rm const.},
\end{eqnarray}
broken only by the anomalous CP-violating terms, i.e. its low-energy effective Lagrangian is parametrized as
\begin{eqnarray}
{\mathcal L}_a =
\frac{1}{2} \partial_\mu a \partial^\mu a
+ \frac{\alpha_s}{4\pi f_a}\, a\, {\rm tr}\, G^{\mu\nu} {\tilde G}_{\mu\nu}+
\frac{s\alpha}{8\pi f_a}\, a\,F^{\mu\nu} {\tilde F}_{\mu\nu} + {\mathcal L}_a^{\rm int} \left[\frac{\partial_\mu a}{f_a};\psi\right],
\label{axion_leff}
\end{eqnarray}
where $s$ is a model dependent parameter, $F$ is the electromagnetic
field strength, and $\psi$ denotes generic Standard Model
fields. The dimensionful axion decay constant $f_a$
determines the strength of the interaction of the axion with the
Standard Model particles. The $\theta$-term in the QCD Lagrangian can then be
eliminated by absorbing it into the axion field, $a=\bar{a} - \bar{\theta}
f_a$. Finally, the topological charge density $\propto \langle
{\rm tr}\, G^{\mu\nu} {\tilde G}_{\mu\nu} \rangle \neq 0$, induced
by topological fluctuations of the gluon fields such as QCD
instantons, provides a nontrivial potential for the axion field $\bar{a}$
which is minimized at zero expectation value, $\langle \bar{a}\rangle =0$:
thus, the $\bar{\theta}$ dependence is wiped out by the axion field,
providing a natural explanation why the electric dipole moment of
the neutron is so small. The nontrivial potential around $\langle \bar{a}\rangle =0$ promotes the elementary particle excitation of the
axion field, the axion, to a pseudo Nambu-Goldstone
boson~\cite{Weinberg:1977ma} (which we will now again denote by $a$) with a non-vanishing, but
parametrically small mass. This mass can be calculated via current
algebra and expressed in terms of the light ($u,d$) quark masses,
the pion mass $m_\pi$ and the pion decay constant $f_\pi$~\cite{Weinberg:1977ma} (cf. also \cite{Amsler:2008zzb}),
      \begin{eqnarray}
m_a =
         \frac{m_\pi f_\pi}{f_a}\frac{\sqrt{m_u m_d}}{m_u+m_d}\simeq { 0.6\,  {\rm meV}}
         \times
         \left(
         \frac{10^{10}\, {\rm GeV}}{f_a}\right) .
         \label{axionmass}
\end{eqnarray}
For large axion decay constant $f_a$, we see that the axion is a prime
example for a WISP~\cite{Kim:1979if}: it is a very weakly interacting (cf. Eq.~(\ref{axion_leff})) sub-eV mass
particle. In particular, its coupling to photons, which for an axion, like for any other pseudo-scalar, should
be of the form,
\begin{eqnarray}
{\cal L}_{a \gamma \gamma} = - \frac{1}{4}\, g\, a\, F_{\mu \nu}\tilde{F}^{\mu \nu} =
g\, a\, \vec{E}\cdot \vec{B} ,
\end{eqnarray}
is very much suppressed~\cite{Bardeen:1977bd},
\begin{eqnarray}
        { g} = \frac{\alpha}{2\pi f_a}
\left( {\frac{2}{3}\,\frac{m_u+4 m_d}{m_u+m_d} - s    }\right)
\sim 10^{-13}\ {\rm GeV}^{-1}          \left(
         \frac{10^{10}\, {\rm GeV}}{f_a}\right).
         \label{axionphotoncoupling}
\end{eqnarray}

Although expected to be small, the guaranteed coupling of axions to
photons, Eq.~(\ref{axionphotoncoupling}), may result, if axions
exist, in observable consequences from processes involving large
electromagnetic fields. These often occur in astrophysical and
cosmological environments (cf. Sect.~\ref{astro}) and can be
prepared in laboratory experiments (cf. Sect.~\ref{searches}).

The proposal of an anomalous Peccei-Quinn shift symmetry is motivated to provide for a solution
of the strong CP problem. This concept has been generalized to other similar WISP candidates
-- axion-like particles (ALPs) -- which may arise as (pseudo) Nambu-Goldstone bosons
from the breaking of other global symmetries such as, for example, family symmetries.
However, in contrast to axions, for generic ALPs a non-zero coupling $g$ to photons is not
guaranteed. Moreover, for them a predictive relation between $g$ and
the mass is missing. Correspondingly, ALP searches, exploiting their interactions with
photons, should try to cover the entire parameter space spanned by $g$ and the mass
of the ALP and not only the restricted parameter space, Eqs.~(\ref{axionmass}) and
(\ref{axionphotoncoupling}), predicted for axions.

\subsubsection{Axions and Axion-Like Particles from String Compactifications\label{Sec:axions_string}}

The existence of axions and ALPs can also be strongly motivated from a top-down
point of view. In fact, when compactifying
the six extra spatial dimensions of string theory they arise quite naturally as Kaluza-Klein zero modes
of antisymmetric tensor fields, which are generically present in all string theories.
Moreover, the (Chern-Simons) couplings of these form fields to the gauge fields,
which are crucially determined by anomaly cancelation conditions, result in the anomalous
CP violating couplings in the low-energy effective Lagrangian, Eq.~(\ref{axion_leff}),
necessary for the solution of the strong CP problem.
Thus, string compactifications suggest plenty of candidates for axions and axion-like
WISPs~\cite{Witten:1984dg,Conlon:2006tq,Svrcek:2006yi,Arvanitaki:2009fg}. However, it is fair to say that they do not
really predict them, because there are several mechanisms known by
which they can be removed from the low-energy spectrum. Only the
ones which escape these mechanisms are WISP candidates.

In the compactification of the weakly coupled heterotic string, a
universal, {\em model-independent} axion appears as the Poincare
dual\footnote{\base To perform this dualization explicitly one
introduces the axion field as a Lagrange multiplier for the Bianchi
identity for $H$, $dH=1/(16\pi^2)({\rm tr}R\wedge R-{\rm tr}F\wedge
F)$, and subsequently integrates over
$H$~\cite{Witten:1984dg,Svrcek:2006yi}.} of the (Neveu-Schwarz)
antisymmetric tensor field $B_{\mu\nu}$, with $\mu$ and $\nu$
tangent to 3+1 dimensional Min\-kow\-ski
space-time~\cite{Witten:1984dg}. Its decay constant $f_a$ is quite
independent of the details of the compactification. To compute it,
one considers the action of an $N=1$ supergravity coupled to an
${\rm E}_8\times {\rm E}_8$ pure gauge theory in 9+1 dimensions,
\begin{equation}
S_{\rm H}
=\frac{2\pi M_s^8}{g_s^2}\int d^{10}x \sqrt{-g}R
-\frac{M_s^6}{2\pi g_s^2}\int\frac{1}{4}{\rm tr} F\wedge\star F
-{\frac{2\pi M_s^4}{g_s^2}\int\frac{1}{2} H\wedge\star H}+\ldots ,
\end{equation}
which describes the dynamics of the massless bosonic excitations of the heterotic string
in terms of the Ricci scalar $R$, the
gauge field strength $F$, and the field strength $H$ of the two-form field $B$.
Compactifying this theory
on a 6 dimensional manifold with volume $V_6$, the
resulting effective action can be matched to its standard normalization in 3+1 dimensions
\begin{equation}
S_{\rm 3+1}
=\frac{M_P^2}{2}\int d^4x\sqrt{-g}\,R
-\frac{1}{4g_{\rm YM}^2}\int d^4x\sqrt{-g}\,{\rm tr}\, F_{\mu\nu} F^{\mu\nu}
-{\frac{1}{f_a^2}\int\frac{1}{2} H\wedge\star H} + \ldots\,,
\end{equation}
with
\begin{eqnarray}
M_P^2 = (4\pi/g_s^2) M_s^8 V_6; \hspace{2ex}
g_{\rm YM}^2 = 4\pi g_s^2/(M_s^{6} V_6); \hspace{2ex}
{ f_a^2=g_s^2/(2\pi M_s^4  V_6)}\,,
\label{heterotic_couplings}
\end{eqnarray}
expressing the reduced Planck mass $M_P=2.4\times 10^{18}$~GeV,
the gauge coupling $g_{\rm YM}$, and the axion decay
constant $f_a$ in terms of the string coupling $g_s$, the string scale $M_s=1/\ell_s$, and the volume
$V_6$. Eliminating the volume $V_6$ and the string scale by means of the first two relations
in Eq.~(\ref{heterotic_couplings}),
one ends up with an axion decay constant of order
of the GUT scale~\cite{Choi:1985je},
\begin{eqnarray}
f_a =\alpha_{\rm YM} M_P/(2\pi \sqrt{2})
\simeq  1.1\times 10^{16}\ {\rm GeV}\,,
{\rm \ for\ } \alpha_{\rm YM}=g^2_{\rm YM}/(4\pi)\sim 1/25\,.
\label{f_a_heterotic}
\end{eqnarray}

{\em Model-dependent} axions arise in the context of weakly coupled
heterotic strings from massless excitations of the two-form
$B$-field on the 6 dimensional compact
manifold~\cite{Witten:1984dg}. Correspondingly, their properties
depend much more on the details of the compactification. Nevertheless, a recent exhaustive study has
elucidated~\cite{Svrcek:2006yi} that also in this case the axion
decay constant cannot be smaller than $10^{15}$~GeV. Similar
conclusions have been drawn for the
axions in strongly coupled heterotic string theory~\cite{Svrcek:2006yi}.
These findings can be easily understood physically: it is the string
scale $M_s$ which mainly determines the axion decay
constant~\cite{Conlon:2006tq}. And in the heterotic case, this scale
is large, e.g. $M_s = \sqrt{\alpha_{\rm YM}/(4\pi)} M_P$, for the
weakly coupled heterotic string (cf.
Eq.~(\ref{heterotic_couplings})).

\begin{figure}[t!]
\centerline{\includegraphics[width=0.9\textwidth]{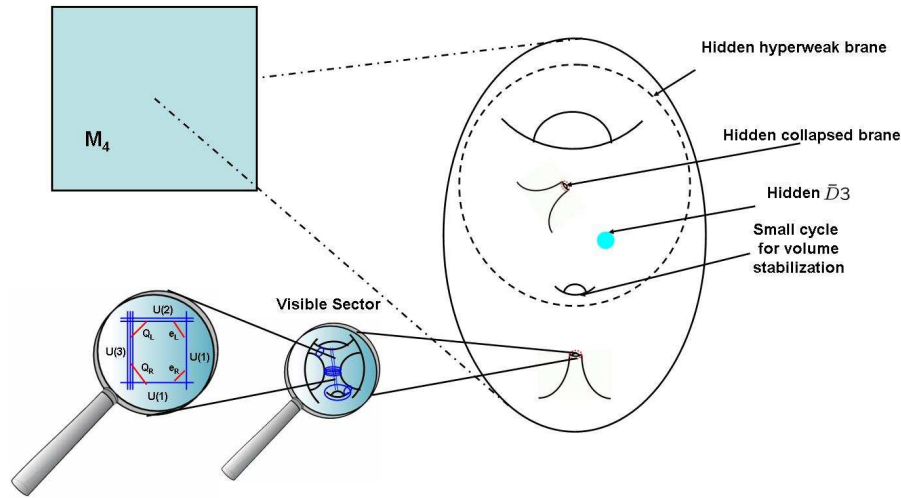}}
\caption{\base In compactifications of type II string theories the
Standard Model is locally realized by a stack of space-time filling D-branes wrapping
topologically non-trivial submanifolds in the compact dimensions. In
general, there can also be hidden sectors localized at different
places. They can arise from branes of different dimension (D3 or D7
branes) which can be either of large extent or localized at
singularities. Light visible and hidden matter particles arise from
strings located at intersection loci and stretching between brane
stacks. }\label{Fig:type_ii_comp}
\end{figure}

This may be different in compactifications of type II string theories which give rise to
``intersecting brane worlds". In these theories, the Standard Model lives on a stack
of D$(3+q)$-branes which are extended along the 3+1 non-compact dimensions and wrap
$q$-dimensional topologically non-trivial submanifolds in the compactification manifold,
while gravity propagates in the bulk, leading to a possibly smaller string scale at the
expense of a larger compactification volume, $M_s\sim g_s M_P/\sqrt{V_6 M_s^6}$ (see Fig.~\ref{Fig:type_ii_comp}).
In type II string theory, the axions come from the massless excitations of the
(Ramond-Ramond) $q$-form gauge field $C_q$. The precise predictions depend
on the particular embedding of the Standard Model, but generically one finds
that the axion decay constant can be substantially
lower than in the heterotic case, varying between~\cite{Conlon:2006tq}
\begin{equation}
f_a\sim \frac{M_P}{\sqrt{V_6 M_s^6}}\sim \frac{M_s}{g_s}\sim 10^{4\div 17}\ {\rm GeV},
\end{equation}
corresponding to a variation of the string scale between the TeV and the GUT scale.

\subsubsection{Scalars and Chameleons}\label{chameleonsect}

Apart from possibly light pseudoscalars, string compactifications generically also predict scalar particles
-- the dilaton and large numbers of moduli -- which appear also massless at the compactification scale.
Essentially massless scalar fields are also often invoked by cosmologists in the context
of dark energy. In fact, a plausible explanation for the apparent acceleration of the cosmic
expansion rate of the universe is provided by the presence of a spatially
homogeneous scalar field which is rolling down a very flat
potential~\cite{Wetterich:1987fm}.

Interactions of very light scalar fields with ordinary matter are strongly
constrained by the non-observation of ``fifth force" effects
leading to {\it e.g.}~violations of the equivalence principle
(cf. Sect.~\ref{Sec:fifth_force}). Correspondingly,
if such particles exist, the forces mediated by them should be either much weaker
than gravity or short-ranged in the laboratory. The latter occurs in theories
where the mass of the scalar field depends effectively on the local
density of matter -- in so-called chameleon field theories~\cite{Khoury:2003aq}.
Depending on the non-linear field self-interactions and on the interactions with the ambient
matter, the chameleon may have a large mass in regions of high density (like the earth),
while it has a small mass in regions of low density (like interstellar space).
Since such particles are able to hide so well from observations and experiments,
they have been named ``chameleons".

\subsection{Ultralight Hidden-Sector Particles\label{Sec:hidden_string}}

Similar to axions and axion-like particles, additional hidden sector U(1) gauge bosons are
also a generic feature arising in string compactifications. These are therefore well motivated
WISP candidates.

In fact, in the standard compactification of the ${\rm E}_8\times {\rm E}_8$ supergravity based
on the heterotic string on a smooth (Calabi-Yau) manifold,
the Standard Model gauge group is embedded in the first ${\rm E}_8$ factor,
whereas the second ${\rm E}_8$ factor comprises a ``hidden gauge group", which interacts with
the first ${\rm E}_8$ factor only gravitationally~\cite{Candelas:1985en}. This second ${\rm E}_8$ factor may be broken in the
course of compactification to products of non-Abelian and U(1) gauge groups.
The occurrence of hidden U(1)s can be studied quite exhaustively in toroidal orbifold compactifications
of the heterotic string, which allow for a systematic scanning of possible gauge group factors and particles
after compactification. Requiring a realistic visible sector, it seems that there are still
a number of models which have possibly massless hidden U(1)s.
In this ``mini-landscape" of orbifold compactifications of the heterotic string~\cite{Lebedev:2008un} one encounters
a breaking of the gauge symmetry to the Standard Model, a hidden sector
non-Abelian gauge symmetry and, typically, at least one hidden U(1), for example, cf.~Ref.~\cite{Lebedev:2009ag},
\begin{equation}
{\rm E}_8\times {\rm E}_8\to
\underbrace{{\rm SU}(3)\times {\rm SU}(2)\times {\rm U}(1)}_{\rm Standard\ Model}
\times
[{\rm SU(6)}\times {\rm U(1)}] .
\end{equation}

Compactifications of type II string theory also suggest the existence of hidden U(1)s,
be it as Kaluza-Klein zero modes of the previously mentioned
(Ramond-Ramond) form fields or as massless excitations of branes.
In fact,
as illustrated in
Fig.~\ref{Fig:type_ii_comp}, type II compactifications generically involve space-time filling
hidden sector branes not intersecting with the Standard Model branes,
often also for global consistency requirements.

Some of these hidden U(1)s may remain unbroken down to very small energy scales.
In this case their dominant interaction with the photon, which is
encoded in the low-energy effective Lagrangian,
\begin{equation}
\mathcal{L} \supset -\frac{1}{4e^2} F_{\mu \nu} F^{\mu \nu}
- \frac{1}{4g_h^2} X_{\mu \nu} X^{\mu \nu}
+ \frac{\chi}{2 e g_h} F_{\mu \nu} X^{\mu \nu}
+ \frac{m_{\gamma^\prime}^2}{2 g^{2}_h} X_{\mu} X^{\mu},
\label{LagKM}
\end{equation}
with $X_\mu$ denoting the hidden U(1) field with field strength $X_{\mu\nu}$ and gauge coupling $g_{h}$,
will be through kinetic mixing~\cite{Holdom:1985ag}, with mixing parameter $\chi$. Therefore, light hidden U(1)s (``hidden photons") are well motivated WISP candidates, since $\chi$ is expected to be small.

In fact, kinetic mixing is generated at one-loop by the exchange of heavy messengers that couple both
to the visible U(1) as well as to the hidden U(1). In the context of compactifications
of the heterotic string, its size has been estimated as~\cite{Dienes:1996zr}
\begin{equation}
\chi \sim \frac{e g_h}{16\pi^2}\,C\,\frac{\Delta m}{M_P}\sim 10^{-5}\div 10^{-17}\,,
{\rm \ for\ } C\gtrsim 10\,,
\end{equation}
where $\Delta m\sim 10^{5\div 17}$~GeV is the possible range of mass splitting in the messenger sector once supersymmetry is broken.

A great variety for possible values of $\chi$ can also be accommodated in type II compactifications
for the mixing between brane-localized hidden U(1)s and the visible U(1). Here, kinetic
mixing can be understood as originating from the exchange of closed strings through the
bulk~\cite{Lust:2003ky,Abel:2008ai}. Generically, one finds~\cite{Goodsell:2009xc}
\begin{equation}
\label{chieq}
\chi \sim \frac{e g_h}{16 \pi^2},
\end{equation}
where the size of the hidden sector gauge coupling $g_h$ depends on the $q$-dimensional
volume ($0\leq q\leq 6$) of the cycle which the hidden brane wraps,
\begin{equation}
g_h^2 \simeq
\frac{2\pi g_s}{V_q M_s^q}
= 2\pi g_s \left( \frac{4\pi}{g_s^2} \frac{M_s^2}{M_P^2}\right)^{q/6}\,,
\label{gsquaredq}
\end{equation}
leading to a quite large range of possible values for the kinetic mixing,
\begin{equation}
10^{-12}\lesssim \chi \lesssim 10^{-3},
\end{equation}
for the string scale varying between a TeV scale and the GUT scale.
Smaller values of kinetic mixing can be obtained in these setups in special cases where the
one-loop contribution is cancelled or vanishes. Moreover,
exponentially suppressed values can be naturally obtained in flux compactifications
with warped throats~\cite{Abel:2008ai}.

Masses for the hidden photons can arise via the standard Higgs mechanism but also via a Stueckelberg mechanism.
In LARGE volume compactifications small, even sub-eV, masses arise quite naturally~\cite{Goodsell:2009xc}.
If the masses arise from a Stueckelberg mechanism, mass and size of the kinetic mixing are typically linked through one scale, the string
scale, and therefore related to each other. Depending on the specific way in which the cycles wrap the singularities one obtains
expressions for the masses like
\begin{equation}
(m^{\rm Stueck}_{\gamma^{\prime}})^2\simeq\frac{g_{s}}{2}\left(\frac{4\pi}{g^2_{s}}\frac{M^{2}_{s}}{M^{2}_{P}}\right)^z,\quad  z=\frac{1}{3},1.
\end{equation}
For example in the case $z=1$ we obtain for a string scale of $M_{s}\sim 1\,{\rm TeV}$, $m_{\gamma^{\prime}}\sim {\rm meV}$ and,  from Eqs.~\eqref{chieq}, \eqref{gsquaredq}, a mixing of
$\chi\sim10^{-12}$.

The predictions for masses arising from the Higgs mechanism are less precise, however, they can also be
tiny, if the supersymmetry breaking scale in the hidden sector is much smaller than in the visible sector.

Occasionally, there is also light hidden matter charged under the hidden U(1)s.
After diagonalization of the gauge kinetic terms by a shift
$X\to X + \chi A$ and a multiplicative hypercharge renormalization,
one observes that the hidden sector particles acquire a minihypercharge
$\epsilon = \chi g_h$~\cite{Holdom:1985ag}. In particular, also an eventual hidden Higgs particle may
be searched for by exploiting its effective minihypercharge~\cite{Ahlers:2008qc}.
In a similar way minicharged particles can also arise from hidden sector magnetic monopoles
if the gauge fields mix via a non-diagonal $\theta$-term~\cite{Bruemmer:2009ky}.
Therefore, light minicharged particles (MCPs) are
also very well motivated WISP candidates.

\section{Astrophysical and Cosmological Constraints on WISPs}\label{astro}

As we have reviewed in the last section, there is a strong physics case for the possible existence of WISPs.
Moreover, their possible masses and couplings span a very wide range
in parameter space. Correspondingly, searches for signatures of WISPs have to exploit a wide variety of
observational and experimental techniques, ranging from cosmology and astrophysics
to terrestrial laboratory experiments. As reviewed in this section, the strongest bounds on the existence of
WISPs presently often come from stellar evolution and cosmology, where to the best of our knowledge observations seem
to agree with the standard budget of elementary particles.
However, there are also some intriguing astronomical observations
which are hard to  explain by  known physics and might be interpreted as indirect hints pointing towards
the existence of WISPs.

\subsection{Bounds from Stellar Evolution}\label{stellar}

Production of WISPs in stars would substantially affect stellar evolution \cite{Raffelt:1996wa}.
WISPs are only rarely produced in the dense plasmas of stellar interiors, but they will easily
escape, contributing directly to the total energy loss of the star.
This has to be contrasted with the standard energy loss of stars, which
occurs mainly due to photons from the stellar surface and neutrinos from the core.
Therefore, the WISP luminosity is enhanced by a huge volume/surface factor, leading to very strong
constraints.

\begin{figure}[t!]
\centerline{\includegraphics[width=.75\textwidth]{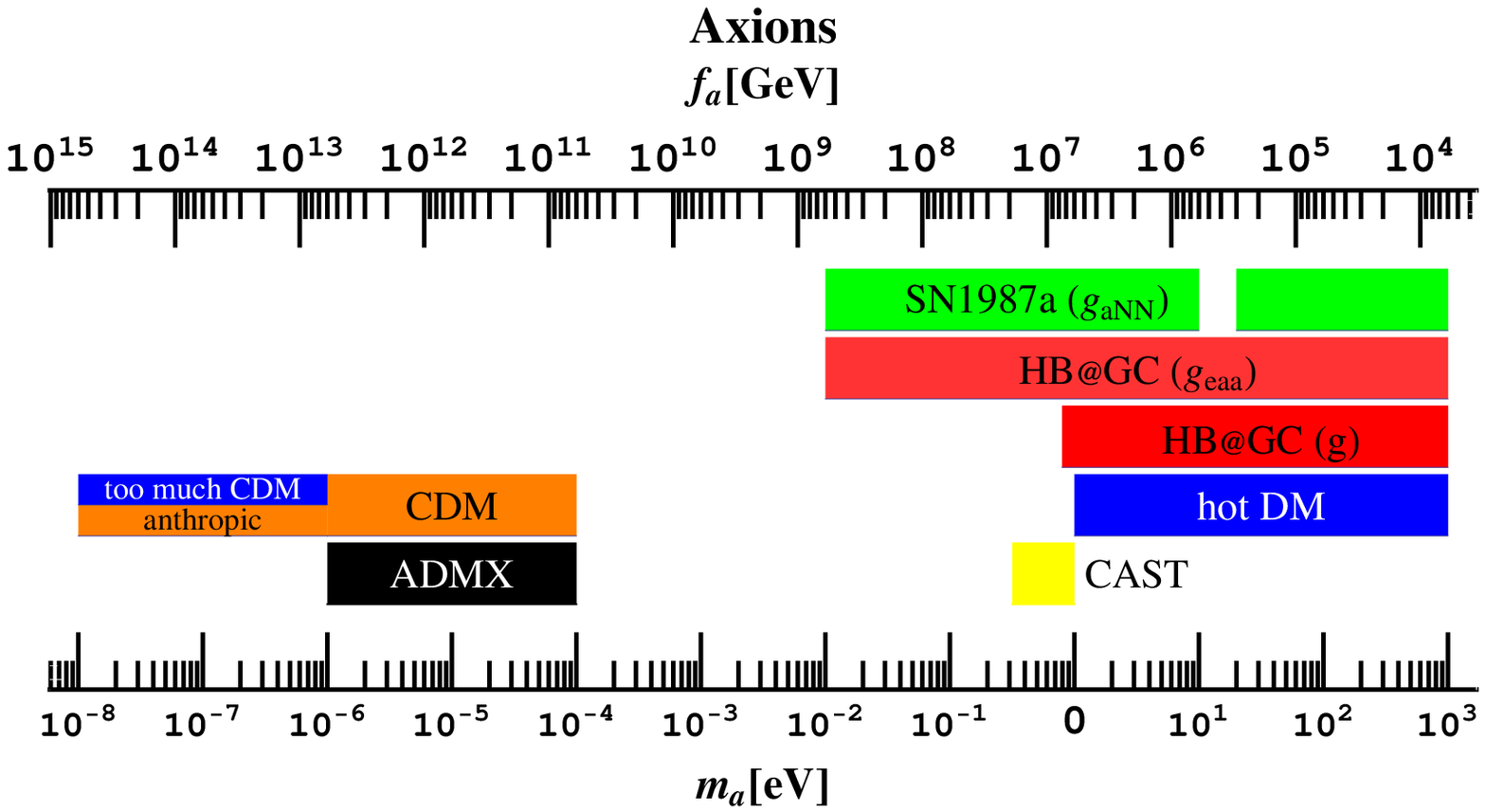}}
\vspace{0.5cm}
\centerline{\includegraphics[width=.6\textwidth]{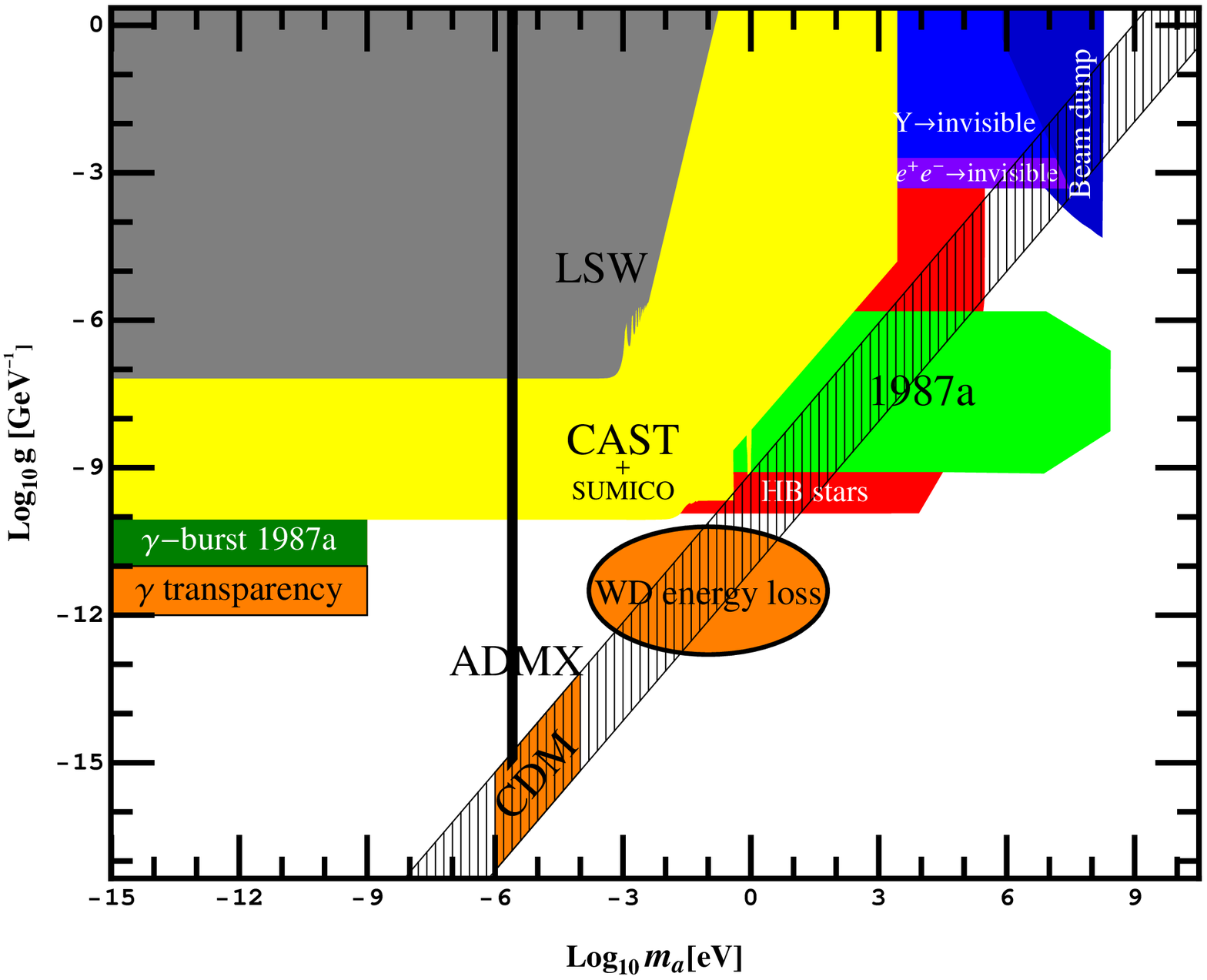}}
\caption{\base Summary of cosmological and astrophysical constraints
for axions (top) (for the mass $m_a$ or decay constant
$f_a$)~\cite{Raffelt:2006cw} and axion-like-particles (bottom) (two
photon coupling $g$ vs. mass $m_{a}$ of the
ALP)~\cite{Andriamonje:2007ew,Schlattl:1998fz,Inoue:2008zp}. See the
text for details. Note that the mass region, where the axion can
be the cold dark matter (the orange regions labeled  ``CDM" in the
plots), can be extended towards smaller masses (larger $f_a\lesssim
10^{16}$\,GeV) by anthropic reasoning. Moreover, in the first plot
the areas marked ``ADMX'' and ``CAST'' show the near future search
ranges. In the second plot the axion band is shown hatched.
We have also marked other
areas with interesting astrophysical hints in orange. For
comparision, we also show laboratory limits from photon regeneration
experiments (ADMX and LSW) as discussed in Section~\ref{searches}.
(Both compilations extended from Ref.~\cite{Redondo:2008en}.) Note
that the limit from ADMX is valid only under the assumption that the
local density of ALPs at earth is given by the dark matter density.
}\label{Fig:axions_astro}
\end{figure}

\begin{figure}[t!]
\centerline{\includegraphics[angle=0,width=1.0\textwidth]{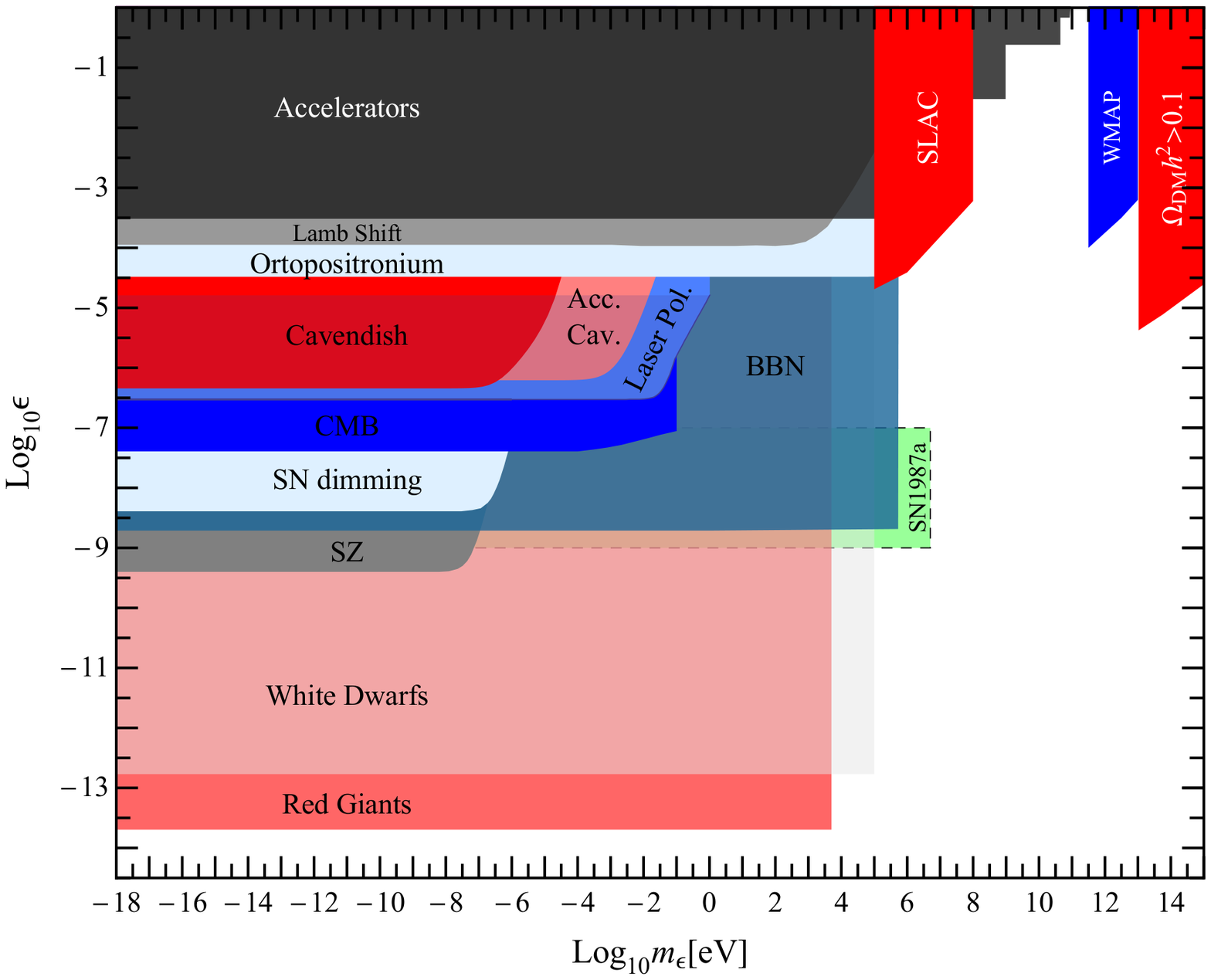}}
\caption{\base
Summary of cosmological and astrophysical constraints for minicharged particles (fractional charge
$\epsilon =Q_\epsilon/e$ vs. mass $m_{\epsilon}$) (compilation from Ref.~\cite{Goodsell:2009xc}).
See the text for details. In addition we also show the laboratory limits discussed in Sect.~\ref{searches}. Moreover, at relatively large
masses and couplings we also have the bounds from accelerator and fixed target experiments (SLAC).}
\label{Fig:mcp_astro}
\end{figure}

Stars evolve fusing increasingly heavier nuclei in their cores. Heavier nuclei require hotter environments, and when a nuclear species is exhausted in the core, the latter slowly contracts and heatens up until it reaches a new burning phase. WISP emission shortens normal burning phases, since the energy loss rate is higher than standard but the total energy is limited by the number of nuclei. On the other hand, WISP emission prolongs intermediate (red giant) phases, since WISP cooling delays reaching the appropriate temperature during the core contraction.
These effects have been used to constrain a variety of WISPs in different stellar environments~\cite{Raffelt:2006cw,Raffelt:1996wa} for which information on evolutionary time scales is  available.
For the standard QCD axions, the best constraints come from white dwarf cooling~\cite{Raffelt:1985nj,Isern:2008nt} through the coupling to electrons and from the duration of the SN1987A neutrino burst~\cite{Raffelt:2006cw} through the coupling
to nucleons (cf. Fig.~\ref{Fig:axions_astro} (top)).
The strongest limits for general ALPs with a two photon coupling and MCPs come from observations of Horizontal Branch (HB) stars in
globular clusters (GC)~\cite{Raffelt:1985nk,Raffelt:1987yu}
(cf. Figs.~\ref{Fig:axions_astro} (bottom) and \ref{Fig:mcp_astro}). For very small masses an even tighter limit on a two-photon coupling of ALPs
can be obtained from the absence of a $\gamma$-ray burst in coincidence with a neutrino burst during the explosion SN 1987a~\cite{Brockway:1996yr}.
The principle behind the
latter bound is that ALPs would be produced in the supernova core from the Primakoff effect and reconverted into $\gamma$-rays inside the galactic magnetic field.

\begin{figure}[t!]
\centerline{\includegraphics[width=1.0\textwidth]{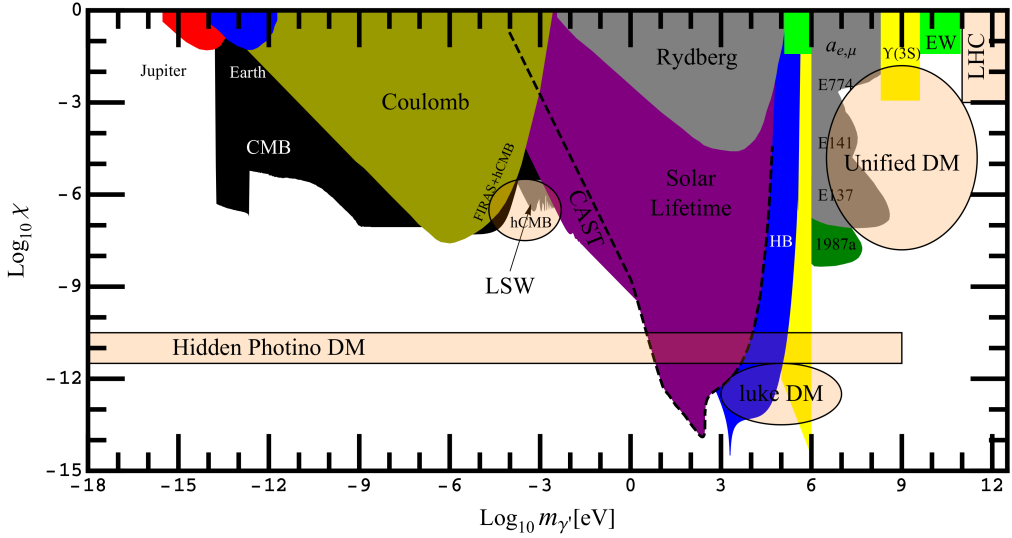}}
\caption{\base
Summary of cosmological and astrophysical constraints for hidden photons (kinetic mixing
$\chi$ vs. mass $m_{\gamma^\prime}$) (compilation from Ref.~\cite{Redondo:priv}).
See the text for details. In addition we also show laboratory limits (see Sect.~\ref{searches} for details on the
constraints in the sub-eV regions; at higher mass we have electroweak precision measurements (EW), bounds from upsilon decays ($\Upsilon_{3S}$)
and fixed target experiments (EXXX)). Areas that are especially interesting are marked in light orange.}
\label{Fig:hp_astro}
\end{figure}

The sun is less sensitive than these other stars to axion or MCP emission, even though its properties are better known. Solar bounds have been obtained from studies of its
lifetime, helioseismology and the neutrino flux~\cite{Schlattl:1998fz,Gondolo:2008dd}, but although the data is more precise the resulting constraints are weaker.
However, as is apparent in Fig.~\ref{Fig:hp_astro}, this is different for hidden photons:
the region in parameter space excluded by the solar lifetime~\cite{Redondo:2008aa} complements
in this case the one excluded by the lifetime of HB stars~\cite{Redondo:2008ec}.

The stellar bounds are very strong but also somewhat vulnerable: they can be considerably relaxed
if the couplings to photons effectively depend on environmental conditions such as the temperature and matter
density~\cite{Jaeckel:2006xm}. This definitely occurs in some specific models, such as the above-mentioned chameleons or in those presented
in Refs.~\cite{Masso:2005ym,Masso:2006gc}.

\subsection{Bounds from Big Bang Nucleosynthesis}

Big Bang Nucleosynthesis (BBN) provides us with a unique probe of
the early universe (for a recent review, see
Ref.~\cite{Iocco:2008va}). At temperatures below $\sim$~MeV, the
weak reactions $p+e^-\leftrightarrow n+ \nu_e$ in the primordial
plasma became ineffective, fixing the neutron/proton density ratio
to $n/p\sim 1/7$. In fact, this ``freeze-out" ratio crucially
depends on the rate of cosmic expansion $H$, which in turn grows with increasing total energy
density $\rho$ of all particles in
the primordial plasma: the larger $\rho$, the sooner the $p$-$n$
freezing the closer n/p becomes to the high temperature value of
1/2. After decoupling, during the proper primordial nucleosynthesis, neutrons are mostly confined into $^4$He nuclei whose
primordial abundance can be measured today, leading to a bound on
the non-standard energy density $\rho_x$ during BBN, usually
expressed as the effective number of extra thermal neutrino species,
\begin{equation}
N_{\nu,x}^{\rm eff}\equiv  \frac{4}{7}\frac{30}{\pi^2 T^4}\rho_x .
\end{equation}
A recent determination of
this number~\cite{Simha:2008zj} resulted in
\begin{equation}
N_{\nu,x}^{\rm eff} = -0.6_{-0.8}^{+0.9},
\end{equation}
for three standard neutrinos.
Therefore, while an extra neutral spin-zero particle thermalized during BBN is allowed,
this is not the case for other WISPs like a mini-charged particle, for which
\begin{equation}
N_{\nu,\rm MCP}^{\rm eff}\geq 1,
\end{equation}
or a massive hidden photon, with
\begin{equation}
N_{\nu,\gamma^\prime}^{\rm eff}=21/16 .
\end{equation}
The interactions of MCPs and $\gamma^\prime$s with the standard bath should not allow thermalization before BBN.
MCPs $\psi$ are produced with a rate $\Gamma(e^+ e^-\to \psi\ \overline \psi)\sim \alpha^2 \epsilon^2 T/2$,  while $\gamma^\prime$s are produced with rate $\Gamma(\gamma e^\pm \to \gamma' e^\pm)\sim \chi^2_{\rm eff}\Gamma_{\rm C}$ with $\Gamma_{\rm C}$ the standard Compton scattering rate. Here $\chi_{\rm eff}$ is the effective $\gamma-\gamma^\prime$ mixing in the plasma, which for sub-eV $\gamma^\prime$ masses is $\chi_{\rm eff}\simeq \chi(m_{\gamma^\prime}/\omega_{\rm P})^2$. The ratio of the $\gamma^\prime$ mass to the plasma frequency, $m_{\gamma^\prime}/\omega_{\rm P}$, is extremely small before BBN so it suppresses $\gamma^\prime$ production with respect to other WISPs.
Correspondingly, one finds, from a comparison
with the expansion rate $H$, that MCPs with $\epsilon < 2\times 10^{-9}$
would be allowed~\cite{Davidson:2000hf} (cf. Fig.~\ref{Fig:mcp_astro}, labelled ``BBN"), but there are no significant bounds for
hidden photons~\cite{Masso:2006gc}.

\subsection{Bounds from the Cosmic Microwave Background\label{Sec:bounds_cmb}}

The cosmic microwave background (CMB) features an almost perfect
blackbody spectrum with ${\cal O}(10^{-5})$ angular anisotropies. It
is released at a temperature $T\sim 0.1$ eV, but the reactions
responsible for the blackbody shape freeze out much earlier, at
$T\sim$ keV. Reactions like $\gamma+...\to$WISP$+...$ would have
depleted photons in a frequency dependent way, which can be
constrained by the precise FIRAS spectrum
measurements~\cite{Fixsen:1996nj}. This can be used to constrain
light MCPs and ALPs~\cite{Melchiorri:2007sq} as well as hidden
photons~\cite{Jaeckel:2008fi}. More generally~\cite{Mirizzi:2009iz},
(resonant) production of hidden photons leads to distortions in the
CMB spectrum measured by FIRAS strongly constraining their existence
in a wide mass range, as can be seen from Fig.~\ref{Fig:hp_astro}
(similar bounds can be obtained for ALPs but they depend on the
unknown strength of the intergalactic magnetic field~
\cite{Mirizzi:2009nq}). Similarly, in presence of MCPs, when the CMB
photons pass through the magnetic field of clusters this leads to a
local distortions of the CMB spectrum in the direction of the
cluster. Such distortions are constrained by
measurements~\cite{Lawrence} of the so-called Sunaev-Zel'dovich (SZ)
effect and lead to strong bounds on
MCPs~\cite{Burrage:2009yz}\footnote{\base Analogously light from distant
supernovae passing through the (less well known) intergalactic magnetic field
would be dimmed by MCP production, again constraining the existence
of such particles~\cite{Ahlers:2009kh} (SN dimming in
Fig.~\ref{Fig:mcp_astro}).}. On the other hand, around $T\sim$ eV
the primordial plasma is so sparse that WISPs would free-stream out
of the density fluctuations, diminishing their contrast. Moreover,
thermal WISPs contribute to the \emph{radiation} energy density,
delaying the matter-radiation equality and reducing the contrast
growth before decoupling. In this respect, they behave identically
to standard neutrinos~\cite{Ichikawa:2008pz}. Therefore, the extra
contribution to the energy density, $\rho_x$ (and the couplings that
would produce it), can again be constrained from the value of
$N_\nu^{\rm eff}$ inferred from analysis of CMB anisotropies and
other large scale structure (LSS) data, e.g. from a recent
analysis~\cite{Simha:2008zj}
\begin{equation}
N^{\rm eff}_{\nu,x} =  -0.1^{+2.0}_{-1.4}
.
\label{cmb_lss_bound}
\end{equation}
This argument has been used to constrain axions~\cite{cosmoaxion}
(cf. Fig.~\ref{Fig:axions_astro} (top), labelled ``hot DM") and meV $\gamma^\prime$s~\cite{Jaeckel:2008fi}
(cf. Fig.~\ref{Fig:hp_astro}, labelled ``FIRAS+hCMB").
However, it should be noted that in the determination of the value of
$N^{\rm eff}_{\nu,x}$ in Eq.~(\ref{cmb_lss_bound}),
Ly-$\alpha$ forest data has been deliberately omitted.
Ly-$\alpha$ has systematically favored values of $N_{\nu,x}^{\rm eff}$ larger than
zero~\cite{Seljak:2006bg}, which could be revealing the existence of a cosmic WISP relic density
(cf. Sec.~\ref{Sec:hidden_cmb}). Alternatively, it may be
due to an incorrect treatment of the bias parameter~\cite{Hamann:2007pi}.

\subsection{Possible Indirect Hints for WISPs\label{hints}}

However, there are also other cosmological and astrophysical puzzles which -- interpreted in terms of WISPs --
may indicate that the latter are just around the corner and that it is of high interest to search for them under
controlled laboratory conditions.

\subsubsection{Axions as Cold Dark Matter}\label{axiondm}

First of all, this concerns the possibility that axions constitute
the cold dark matter (CDM) in the universe. In fact, for very weak
coupling, i.e. large decay constant $f_a$, the ultra-light axions
are produced non-thermally in the early universe.  At early times,
at temperatures well above the QCD phase transition, the axion is
effectively massless and the corresponding field can take any value,
parameterized by the ``misalignment angle" $\theta_i$. Later, as the
temperature of the primordial plasma falls below the hadronic scale,
$T\lesssim $~GeV, the axion develops its mass $m_a$ due to
non-perturbative (topological instanton) effects. When the mass
$m_a$ becomes of order the Hubble expansion rate, the axion field
will start to oscillate around its mean value $\langle a\rangle =0$.
These coherent and spatially uniform oscillations correspond to a
coherent state of non-relativistic axion particles, whose
contribution to today's energy density, in terms of the critical
energy density, can be estimated\footnote{\base It should be noted that
this is a relatively crude estimate. In principle the amount of axions also depends
on the order in which cosmological events took place. In particular, whether the breaking of the Peccei-Quinn symmetry
occurred before or after inflation. In the latter case, for example, there are additional contributions
from the formation and decay of cosmic strings and domain walls. Typically, these contributions
are of similar order to the axion density produced in the misalignment mechanism~\cite{Sikivie:2006ni}.} as~\cite{Preskill:1982cy}
\begin{equation}
\Omega_a h^2 = \kappa_a \left( \frac{f_a}{10^{12}\ {\rm GeV}}\right)^{1.175} \theta_i^2,
\end{equation}
where $0.5\lesssim \kappa_a\lesssim $~few. Therefore, for generic
values of the misalignment angle, $\theta_i=\mathcal{O}(1)$, the
axion could be the main constituent of CDM in the universe,
$\Omega_{\rm CDM} h^2 \sim 0.1$, if its decay constant is of order
$f_a\sim 10^{12}$~GeV, corresponding to a mass in the $10\,\mu$eV range.
Larger values of $f_a$ (cf. Fig.~\ref{Fig:axions_astro} (top);
labeled ``CDM") are excluded because they would lead to an
overclosure of the universe. However, if the initial $\theta_i$ is small,
values of $f_a$ near the GUT or Planck scale are still possible.
This could be due to a finetuning but also due to anthropic
selection~\cite{Hertzberg:2008wr}.

Axion dark matter may indeed explain two additional puzzling
observations: firstly there is an interesting alignment in the multipoles of the CMB \linebreak anisotropies~\cite{Tegmark:2003ve}
and secondly the rotational curves of a number of galaxies provides
evidence for an additional structure in the galactic halo, so-called caustic rings~\cite{Kinney:1999rk}.
Both observations may be explained by Bose-Einstein condensation of dark matter axions~\cite{Sikivie:2009fv}.

Finally, in the context of dark matter it should be noted that axions could also contribute a
hot dark matter component. As for neutrinos the fraction of this hot dark matter is proportional to the mass of
the axion. Recent constraints on the size of a possible hot dark matter component provide a
bound of $m_{a}\lesssim 1.2\,{\rm eV}$ (see Fig.~\ref{Fig:axions_astro} (top))~\cite{cosmoaxion}.

\subsubsection{Non-Standard Energy Loss in White Dwarfs}

A possible non-standard energy loss has been recently identified in the white dwarf luminosity function~\cite{Isern:2008nt}.
As pointed out by the authors this is compatible with the existence of axions with an axion-electron coupling,
\begin{equation}
g_{eea}\simeq 10^{-13},
\end{equation}
suggesting an axion decay constant and axion mass\footnote{\base For the mass of an ALP actually everything below
$\sim$ keV would be acceptable.} of
\begin{equation}
f_a\sim g_{eea} m_e \sim  {\rm few}\times 10^{9}\ {\rm GeV},\hspace{6ex}
m_a\sim {\rm meV},
\label{fa_benchmark_axion_wd}
\end{equation}
respectively.
In typical models, the coupling to photons is then
\begin{equation}
g\sim \alpha/f_a  \sim 10^{-12}\ {\rm GeV}^{-1},
\label{g_benchmark_axion_wd}
\end{equation}
i.e. very close to the stellar evolution bounds.
But, apart from the obvious possibility that
a more conventional explanation may be found for this non-standard energy loss, it should
be noted that this could also be explained in terms of hidden photons coupling via
non-renormalizable operators to electrons, cf. Ref.~\cite{Hoffmann:1987et}.
Nevertheless, the values in Eqs.~(\ref{fa_benchmark_axion_wd}) and
(\ref{g_benchmark_axion_wd}) should be considered as benchmark values for
future axion searches.  Fortunately, as we will see in the next section, this region of parameter space
is not completely off the possibilities to be checked in a controlled laboratory experiment
based on photon regeneration.

\subsubsection{Hints for Cosmic Photon Regeneration\label{Sec:hints_cosmic}}

It has been argued that recent observations in TeV gamma astronomy may point towards the
existence of ALPs or at least will help to give sensible new constraints on their existence.
Quite distant astrophysical sources have been observed in TeV gamma rays by H.E.S.S. and
MAGIC. This appeared to be quite puzzling, since the gamma ray absorption rate
due to electron/positron pair production off the extragalactic background light (EBL)
was believed to be too strong to allow
for their observation~\cite{Aharonian:2005gh,Mazin:2007pn}. Clearly, a conventional explanation is either
that the EBL is less dense than expected and/or that the source
spectra are harder as previously thought. Alternatively,
such a high transparency of the universe may also be explained by the conversion of
gamma rays into ALPs in the magnetic fields around the gamma ray sources. These ALPs would travel then unimpeded
until they reach our galaxy and reconvert into photons in the galactic magnetic
fields~\cite{Hochmuth:2007hk}. Alternatively, the conversion/reconversion could take place in
the intergalactic magnetic fields~\cite{DeAngelis:2007dy} (see Ref.~\cite{Mirizzi:2009aj}
for a comprehensive bibliography and the current status).
The intergalactic magnetic fields are not well known but the assumption of being organized in randomly oriented patches would produce
a characteristic scatter in luminosity relations.
A powerful signature of cosmic photon regeneration could therefore emerge if the reconstructed
EBL along different lines of sight towards different TeV gamma sources were to display such a
characteristic scatter~\cite{Mirizzi:2009aj}.

ALPs may also leave their imprints in luminosity relations of active galactic nuclei.
In fact, mixing between photons and ALPs in the random magnetic fields in galaxy clusters
may induce a characteristic scatter in the relations of X-ray vs. optical luminosities of compact sources
in these clusters. Evidence for such an effect has recently been found in an analysis of
luminosity relations of about two hundred active galactic nuclei, providing
a strong hint for the possible existence of a very light axion-like particle~\cite{Burrage:2009mj}.

Furthermore, photon - ALP mixing could also explain the puzzling origin of the (debated)
correlation of arrival directions of ultra high energy cosmic rays and BL-Lac objects observed by AGASA and
HiRes~\cite{Fairbairn:2009zi} (see, however, Ref.~\cite{Albuquerque:2010rq}).
Moreover, the observed alignment of the polarization vectors of very distant quasars
may also be explained by selective photon disappearance from photon-ALP oscillations~\cite{Payez:2008pm}.
Finally, it is also worth nothing that very similar ALPs have been invoked to solve some problematic aspects of the X-ray activity of the Sun,
the longstanding corona problem and the triggering of solar flares~\cite{Zioutas:2008ie}.

Intriguingly, in all the above mentioned hints for cosmic photon regeneration, the required ALP should
have a very small mass, say~\cite{Burrage:2009mj}
\begin{equation}
m_{a}\ll 10^{-12}\,{\rm eV}\div 10^{-9}\,{\rm eV},
\label{mass_benchmark_ALP_transp}
\end{equation}
and a coupling in the range
\begin{equation}
g\sim 10^{-12}\div 10^{-11}\ {\rm GeV}^{-1}.
\label{g_benchmark_ALP_transp}
\end{equation}
Again, these values should be considered as important benchmarks for future
laboratory searches for ALPs. As we will see, photon regeneration experiments can do a very good
job here.

\subsubsection{Dark matter from hidden photons\label{Sec:dm_hidden}}
Dark matter could also be directly related to hidden photons.
Indeed there are several different way in which a hidden U(1) could contribute to dark matter.

First of all in the region $(\chi,m_{\gamma^\prime} )\sim (10^{-12}, 0.1\ {\rm MeV}$), labeled ``luke DM" in
Fig.~\ref{Fig:hp_astro}, the hidden photon itself could be a lukewarm dark matter candidate~\cite{Pospelov:2007mp,Redondo:2008ec}.

For somewhat larger mixing angles $(\chi,m_{\gamma^\prime} )\sim (10^{-11}, \lesssim  100\,{\rm GeV}$), in the region labeled ``Hidden
Photino DM" in Fig.~\ref{Fig:hp_astro}, the supersymmetric partner
of the hidden photon, the hidden photino, is a promising dark matter candidate, if its mass is
in the 10 to 150 GeV range~\cite{Ibarra:2008kn}.
Moreover, for $(\chi,m_{\gamma^\prime} )\sim (10^{-23}, 0$), the hidden photino, with mass in the TeV range, could be a candidate of decaying dark matter, giving rise to
the above mentioned excesses observed in galactic cosmic
ray positrons and electrons~\cite{Shirai:2009kh}.

Finally, in the region $(\chi,m_{\gamma^\prime} )\sim (10^{-4},  {\rm GeV}$), labeled ``Unified DM" in
Fig.~\ref{Fig:hp_astro} the hidden photon
plays an important role in models where the dark matter resides in the
hidden sector~\cite{ArkaniHamed:2008qn}. These models aim at a unified description of unexpected observations
in astroparticle physics, notably the positron excess observed by
the satellite experiment PAMELA \cite{Adriani:2008zr} and the annual modulation signal seen by the direct dark matter search experiment DAMA~\cite{Bernabei:2008yi}.
The massive hidden U(1) can then mediate ``Dark Forces''.
These values are also accessible to accelerator searches~\cite{ArkaniHamed:2008qp,Baumgart:2009tn}
and have been motivated in various supersymmetric scenarios~\cite{Chun:2008by,Baumgart:2009tn}
(see also Ref.~\cite{Suematsu:2006wh}).

\subsubsection{A hidden CMB?\label{Sec:hidden_cmb}}
As discussed in Sect.~\ref{Sec:bounds_cmb}, hidden photons contribute to the effective number of neutrinos.
Indeed, such a contribution would lead to a higher number of effective neutrinos at the time of CMB and large scale structure
formation than at the time of BBN. Interestingly, some global cosmological analyses
that take into account precision cosmological
data on the cosmic microwave background and on the large scale structure of the universe
appear to require some extra radiation energy density from invisible particles apart from the three known neutrino
species. The case for this was strengthened by the recently released WMAP 7 year data~\cite{Komatsu:2010fb}.
Hidden photons in the parameter region $(\chi,m_{\gamma^\prime} )\sim (10^{-6}, 0.2\ {\rm meV}$), labeled ``hCMB" in
Fig.~\ref{Fig:hp_astro}, lead to a natural explanation of this finding~\cite{Jaeckel:2008fi}.

As a side remark we note that these parameter values also allow for interesting technological
applications of hidden photons~\cite{Jaeckel:2009wm}.

\section{WISP Searches with Low-Energy Photons\label{searches}}
One of the most striking features of many new light bosons is that
one can have photon -- light boson oscillations in very much the
same way as the different neutrino species oscillate into each
other. Below we will start with the description of so-called ``light
shining through a wall'' (LSW) experiments, which most directly make use of
this oscillation phenomenon. We will also use this opportunity to
introduce the basic equations governing these oscillation phenomena.

\subsection{Photon Regeneration Experiments}

\subsubsection{Light Shining Through a Wall -- Theory}
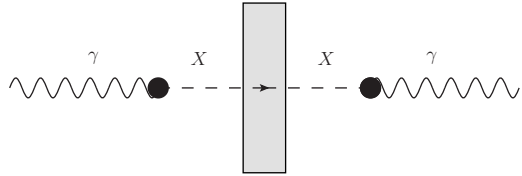
\begin{figure}[t!]
\begin{center}
  \scalebox{0.5}[0.5]{
  \begin{picture}(386,130) (159,-207)
    \SetWidth{1.0}
    \SetColor{Black}
    \GBox(336,-206)(368,-78){0.882}
    \Line[dash,dashsize=8.4,arrow,arrowpos=0.5,arrowlength=5,arrowwidth=2,arrowinset=0.2](272,-142)(432,-142)
    \Photon(160,-142)(272,-142){7.5}{6}
    \Photon(432,-142)(544,-142){7.5}{6}
    \Vertex(432,-142){8}
    \Vertex(432,-142){8}
    \Vertex(272,-142){8}
    \SetOffset(0,-10)
    \Text(224,-110)[c]{\Large{\Black{$\gamma$}}}
    \Text(480,-110)[c]{\Large{\Black{$\gamma$}}}
    \Text(304,-110)[c]{\Large{\Black{$X$}}}
    \Text(400,-110)[c]{\Large{\Black{$X$}}}
  \end{picture}
  }
  \end{center}
\caption{\base
Schematic of a ``light-shining-through a wall'' experiment.
An incoming photon $\gamma$ is converted into a new particle $X$
which interacts only very weakly with the opaque wall. It passes
through the wall and is subsequently reconverted into an ordinary
photon which can be detected.
}\label{Fig:lsw}
\end{figure}

One of the most striking consequences of the photon -- light boson
oscillations is the possibility of ``light shining through a wall''~\cite{Okun:1982xi,Anselm:1986gz,VanBibber:1987rq}.
This is exploited in experiments of the same name. A schematic setup
is shown in Fig.~\ref{Fig:lsw}. The idea is as follows. If an
incoming photon is somehow converted into a WISP the latter can
transverse an opaque wall without being stopped. On the other side
of the wall the WISP could then reconvert into a photon.

\begin{figure}[t!]
\begin{center}
\subfigure[]{\scalebox{0.3}[0.3]{
  \begin{picture}(322,141) (95,-63)
    \SetWidth{1.0}
    \SetColor{Black}
    \Photon(96,13)(176,13){7.5}{4}
    \GBox(240,-51)(272,77){0.882}
    \Line[dash,dashsize=2,arrow,arrowpos=0.5,arrowlength=5,arrowwidth=2,arrowinset=0.2](176,13)(336,13)
    \Photon(176,13)(176,-51){7.5}{3}
    \COval(176,-51)(11.314,11.314)(45.0){Black}{White}\Line(170.343,-56.657)(181.657,-45.343)\Line(170.343,-45.343)(181.657,-56.657)
    \Photon(336,13)(336,-51){7.5}{3}
    \Photon(336,13)(416,13){7.5}{4}
    \COval(336,-51)(11.314,11.314)(45.0){Black}{White}\Line(330.343,-56.657)(341.657,-45.343)\Line(330.343,-45.343)(341.657,-56.657)
  \end{picture}
  }
\label{zooalp}}
  \hspace*{0.2cm}
\subfigure[]{\scalebox{0.3}[0.3]{
  \begin{picture}(322,130) (95,-73)
    \SetWidth{1.0}
    \SetColor{Black}
    \Photon(96,2)(176,2){7.5}{4}
    \GBox(240,-62)(272,66){0.882}
    \Photon(336,2)(416,2){7.5}{4}
    \ZigZag(176,2)(336,2){7.5}{8}
    \SetWidth{3.0}
    \Line(160.002,17.998)(191.998,-13.998)\Line(191.998,17.998)(160.002,-13.998)
    \Line(320.002,17.998)(351.998,-13.998)\Line(351.998,17.998)(320.002,-13.998)
  \end{picture}
  }
\label{zoohp} } \hspace*{0.2cm}
 \subfigure[]{\scalebox{0.3}[0.3]{
  \begin{picture}(386,141) (63,-63)
    \SetWidth{1.0}
    \SetColor{Black}
    \GBox(240,-51)(272,77){0.882}
    \Photon(128,13)(64,13){7.5}{3}
    \Arc[arrow,arrowpos=0.5,arrowlength=5,arrowwidth=2,arrowinset=0.2](160,13)(32,270,630)
    \ZigZag(192,13)(320,13){7.5}{6}
    \Arc[arrow,arrowpos=0.5,arrowlength=5,arrowwidth=2,arrowinset=0.2](352,13)(32,270,630)
    \Photon(384,13)(448,13){7.5}{3}
    \Photon(146,-16)(127,-52){7.5}{2}
    \Photon(176,-15)(192,-51){7.5}{2}
    \Photon(336,-15)(320,-51){7.5}{2}
    \Photon(368,-14)(384,-52){7.5}{2}
    \COval(128,-51)(11.314,11.314)(45.0){Black}{White}\Line(122.343,-56.657)(133.657,-45.343)\Line(122.343,-45.343)(133.657,-56.657)
    \COval(192,-51)(11.314,11.314)(45.0){Black}{White}\Line(186.343,-56.657)(197.657,-45.343)\Line(186.343,-45.343)(197.657,-56.657)
    \COval(320,-51)(11.314,11.314)(45.0){Black}{White}\Line(314.343,-56.657)(325.657,-45.343)\Line(314.343,-45.343)(325.657,-56.657)
    \COval(384,-51)(11.314,11.314)(45.0){Black}{White}\Line(378.343,-56.657)(389.657,-45.343)\Line(378.343,-45.343)(389.657,-56.657)
  \end{picture}
  }
\label{zoomcp}}
  \end{center}
\caption{\base
Explicit processes contributing to LSW for various WISPs.
From left to right we have photon -- ALP, photon -- hidden photon
and photon -- hidden photon oscillations facilitated by
MCPs.}\label{Fig:zoo}
\end{figure}
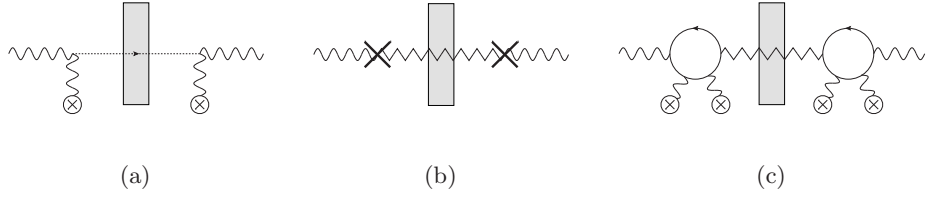

This type of experiment is sensitive to a whole variety of WISPs as shown in Fig.~\ref{Fig:zoo}.
In particular, the classic axion or axion-like particles can be searched for by
employing a magnetic field in the conversion regions.
This facilitates the conversions of photons into
axions via the two photon interaction predicted for axion-like
particles (cf. Fig.~\ref{zooalp}).

These oscillation phenomena can be described using a non-diagonal
mass term in the equations of motion decribing the photon $A$
(with energy $\omega$) and
the new particle $X$ (cf. also Ref.~\cite{Raffelt:1987im}),
\begin{equation}
\label{eom}
[\omega^2\mathbf{1}+\partial^{2}_{z}\mathbf{1}-{\mathcal{M}}^{X}]
\left(
\begin{array}{c}
                               A \\
                               X \\
                             \end{array}
                           \right)=0,
\end{equation}
where we have suppressed the Lorentz structure. Indeed, for the
types of particles discussed in Sect.~\ref{physicscase} the
equations of motion always separate into the two possible linear
polarizations but the mass matrix may differ for the different
polarization directions.

The solutions to the equations of motion are of the form,
\begin{equation}
v_{1}=\exp(-\mathbf{i}(\omega t-k_{1}z))\left(
                                         \begin{array}{c}
                                           1 \\
                                           \delta \\
                                         \end{array}
                                       \right),
                                       \quad
v_{2}=\exp(-\mathbf{i}(\omega t-k_{2}z))\left(
                                         \begin{array}{c}
                                           -\delta \\
                                           1 \\
                                         \end{array}
                                       \right).
\end{equation}
If the off-diagonal entry in the mass matrix is small we can obtain
simple analytical formulas for the mixing angle,
\begin{equation}
\tan(2\,\delta)=2\frac{\mass^{X}_{12}}{\mass^{X}_{11}-\mass^{X}_{22}},
\end{equation}
and the wave numbers for the two mass eigenstates,
\begin{equation}
k^2_{1}=\omega^2-\mass^{X}_{11},\quad
k^{2}_{2}=\omega^{2}-\mass^{X}_{22}.
\end{equation}
Using these it is straightforward to find the transition amplitudes,
\begin{equation}
\label{amplitude} A(\gamma\to
X)=\delta\left[\exp(\mathbf{i}k_{1}z)-\exp(\mathbf{i}k_{2}z)\right],
\end{equation}
from which we can obtain
\begin{eqnarray}
\label{transition} P(\gamma\to X,\ell)\!\!&=&\!\!P(X\to\gamma,\ell)=|A(\gamma\to X)|^2
\\\nonumber
\!\!&=&\!\!|\delta|^2[\exp(-2\Imag(k_{1})\ell)+\exp(-2\Imag(k_{2})\ell)
\\\nonumber
&&\quad\quad\quad\quad\quad\quad\quad\quad\quad-2\exp(-\Imag(k_{1}+k_{2})\ell)\cos(\Real(k_{1}+k_{2})\ell)].
\end{eqnarray}

In a light shining through a wall experiment the photon must convert
into a WISP and back. Therefore the total probability for a photon
to arrive at the detector is\footnote{\base Here, we neglect the typically
very small absorption probability of the WISP $X$ inside the wall.},
\begin{equation}
\label{lswprob} P(\gamma\to X,\ell_{1})P(X\to\gamma,\ell_{2}).
\end{equation}
This probability can be enhanced by using optical cavities to
reflect the light back and forth inside the production and
regeneration regions~\cite{Hoogeveen:1990vq,Sikivie:2007qm}. In the production region one can easily imagine that
the probability is enhanced by the number of passes towards the
wall. On the regeneration side such an enhancement is not as
obvious. Nevertheless, a \emph{resonant} cavity on the regeneration
side allows for an additional enhancement corresponding to the
number of passes inside this cavity as well. The probability
including these improvements is therefore,
\begin{equation}
\label{enhanced} P_{\rm LSW}=N_{1}N_{2}P(\gamma\to
X,\ell_{1})P(X\to\gamma,\ell_{2}),
\end{equation}
where $N_1$ ($N_2$) is the effective number of passes in the respective cavity
divided by a factor two.

Let us now consider the three sets of particle discussed in Sect.~3:
axion-like particles, massive hidden photons, and massless hidden
photons with additional minicharged particles.

Axion-like particles
couple to two photons. In this case the photon -- axion-like
particle oscillations have to be facilitated by the presence of a
magnetic field as shown in Fig.~\ref{Fig:lsw} (a) which provides for
one of the two photons in the interaction. The presence of the
magnetic field marks a preferred direction. Therefore the matrix
$\mass$ can now depend on the polarization direction. Indeed one
finds,
\begin{eqnarray}
\mass^{a^{-}}_{\parallel}\!\!&=&\!\!\left(
                            \begin{array}{cc}
                              0 & -gB\omega \\
                              -gB\omega & m^{2}_{a} \\
                            \end{array}
                          \right)
,\quad\quad\mass^{a^{-}}_{\perp}=0,
\\\nonumber
\mass^{a^{+}}_{\parallel}\!\!&=&\!\!0,\quad\quad\quad\quad\quad\quad\quad\quad\quad\,\,
\mass^{a^{+}}_{\perp}=\left(
                            \begin{array}{cc}
                              0 & -gB\omega \\
                              -gB\omega & m^{2}_{a} \\
                            \end{array}
                          \right),
\end{eqnarray}
where $a^{-}$ indicates a pseudo-scalar axion-like particle coupling
to $F\tilde{F}$ and $a^{+}$ is a scalar coupling to $F^2$. Moreover,
the subscripts $\parallel,\perp$ indicate the polarization direction
with respect to the magnetic field.
Since the mass matrix is real
the expression for the probability in an LSW experiment, Eqs.~\eqref{transition},
\eqref{lswprob}, simplifies. If the polarization of the laser is at an
angle $\theta$ with respect to the magnetic field, the probability
for a pseudoscalar reads,
\begin{equation}
\label{axionprob}
P_{\rm LSW}=16\frac{(gB\omega\cos(\theta))^4}{m^{8}_{a}}
\sin^2\left(\frac{\ell_{1}m^2_{a}}{4\omega}\right)
\sin^2\left(\frac{\ell_{2}m^{2}_{a}}{4\omega}\right).
\end{equation}
For a scalar the $\cos(\theta)$ has to be replaced by a
$\sin(\theta)$. Moreover, for simplicity we have made the additional
assumption  $\omega\gg m_{a}$.

The transition into hidden photons occurs also in absence of a
magnetic field. Accordingly there is no preferred direction and both
polarizations have the same mass-mixing,
\begin{equation}
\label{hiddenmass}
\mass^{\gamma^{\prime}}=m_{\gamma^\prime}^2\left(
               \begin{array}{cc}
                 \chi^2 & -\chi \\
                 -\chi & 1 \\
               \end{array}
             \right).
\end{equation}
The corresponding transitions are depicted in Fig.~\ref{Fig:lsw} (b) and
the transition probability is
\begin{equation}
\label{hpprob}
P_{\rm LSW}=16\chi^4\sin^2\left(\frac{\ell_{1}m_{\gamma^\prime}^2}{4\omega}\right)
\sin^2\left(\frac{\ell_{2}m_{\gamma^\prime}^2}{4\omega}\right).
\end{equation}

Finally, for the combination of massless hidden photons and
minicharged particles we again need a magnetic field to allow for a
transition,
\begin{equation}
\mass^{\gamma^{\prime}+{\rm MCP}}= -2\omega^2{e^{2}_{{h}}}\Delta
N_{i}\left(
  \begin{array}{cc}
    +\chi^2  & -\chi  \\
    -\chi  & +1 \\
  \end{array}
\right).
\end{equation}
Here, $i=\parallel, \perp$ again
indicate the polarization with
respect to the magnetic field and $\Delta N_{i}$ are the magnetic field dependent, complex refractive indices describing
the refraction and absorption due to the virtual and real production of MCPs (see \cite{Ahlers:2007rd} for details).

\begin{figure}[t!]
\subfigure[]{\includegraphics[width=.5\textwidth]{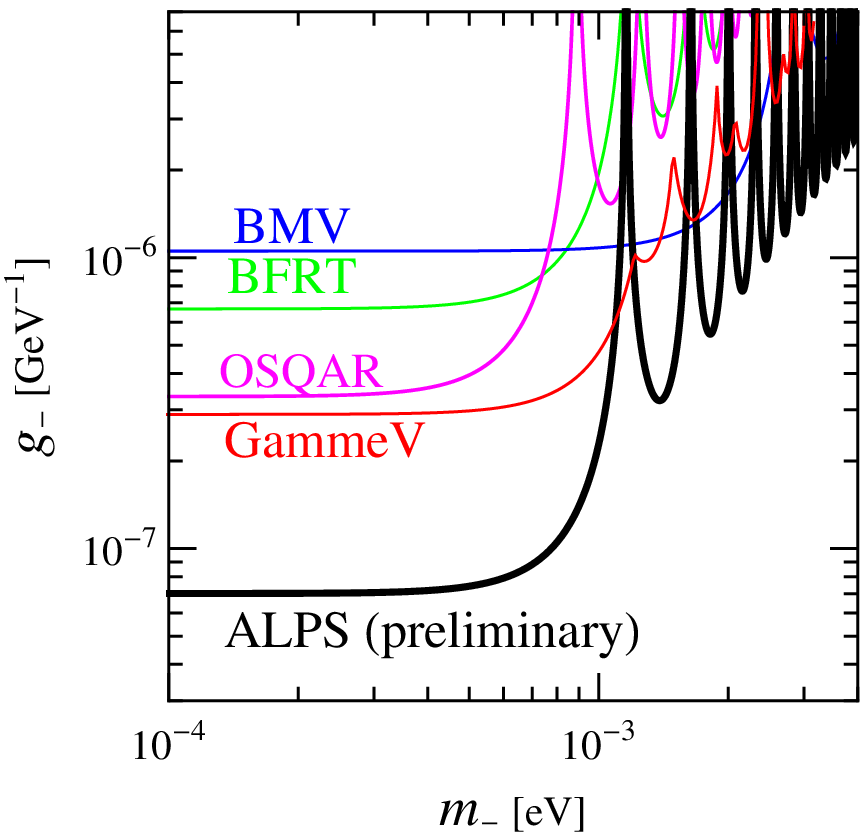}}
\hfill
\subfigure[]{\includegraphics[width=.5\textwidth]{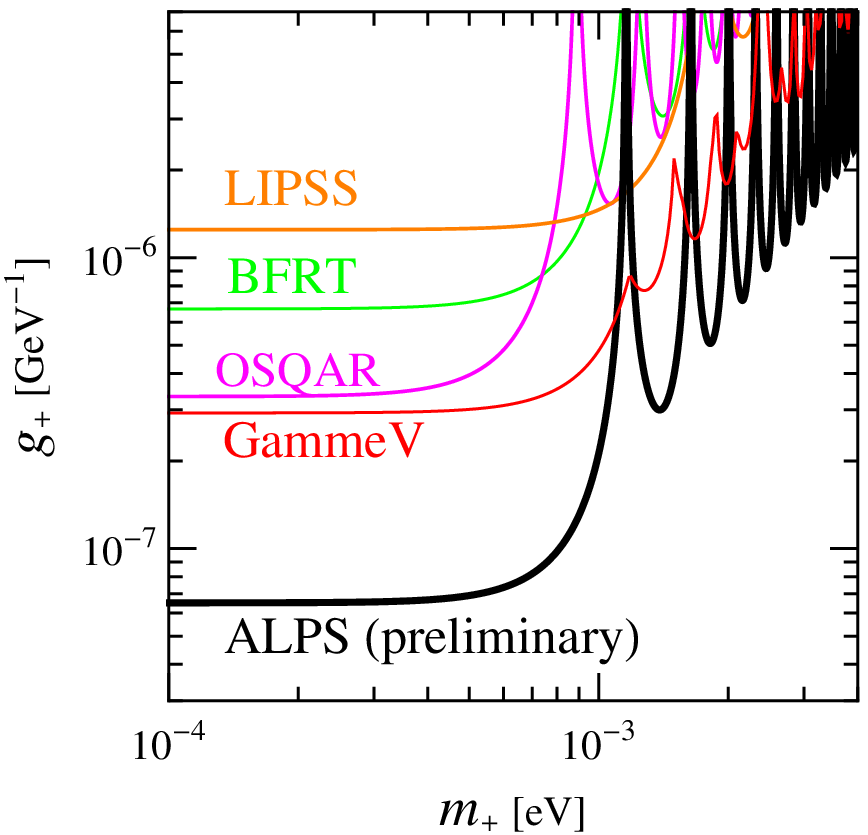}}
\subfigure[]{\includegraphics[width=.5\textwidth]{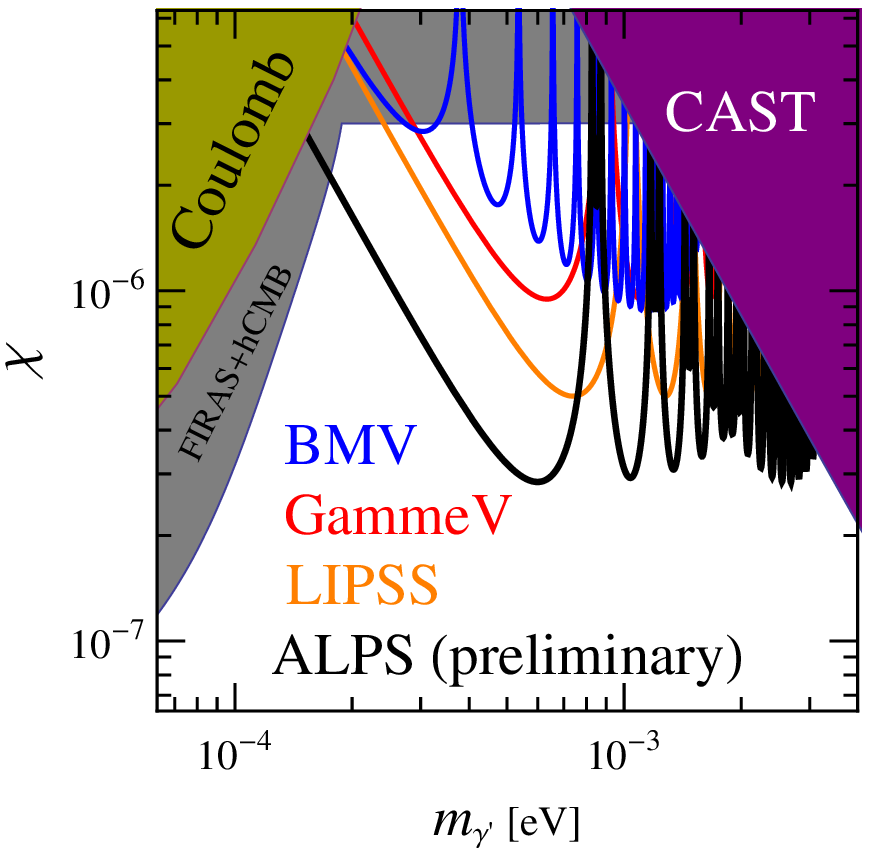}}
\hfill
\subfigure[]{\includegraphics[width=.5\textwidth]{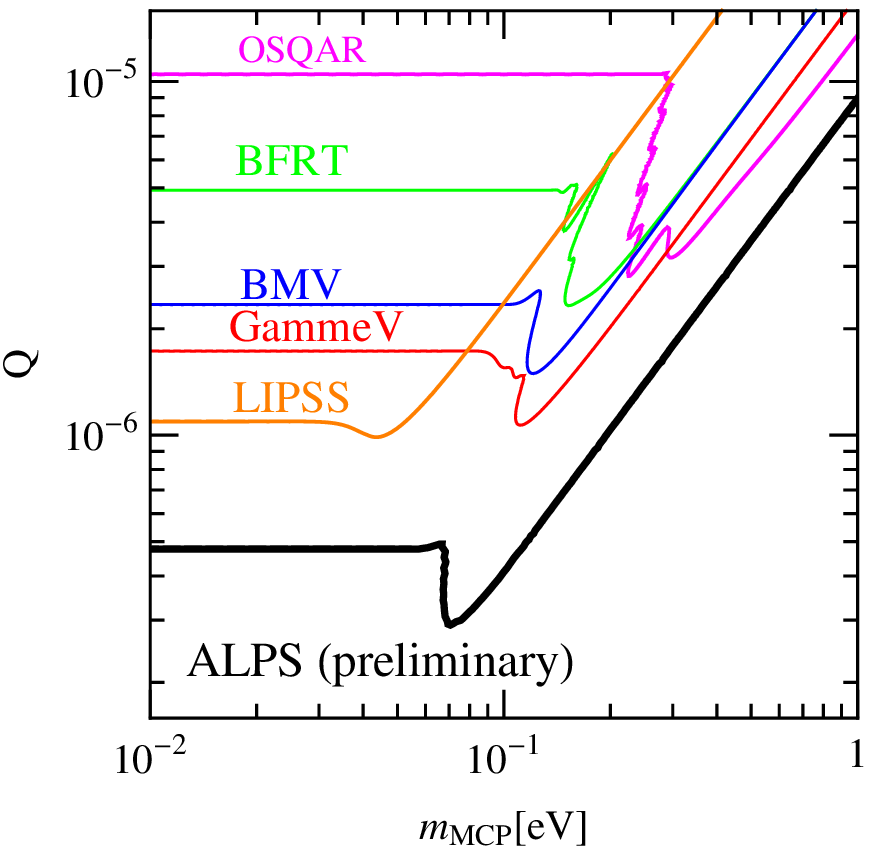}}
\caption{\base
Sensitivities of LSW experiments. Top panels:
pseudoscalar (left) and scalar (right) axion-like particles.
The results from ALPS are preliminary.
Bottom panels: massive hidden photons (left) and massless hidden
photons with an additional minicharged particle (right).
Compilation from Ref.~\cite{ALPS}.}\label{lswresult}
\end{figure}

\subsubsection{Light Shining Through a Wall - Experiments}\label{lswexperiments}

As the name suggests most of these experiments employ laser light in
the optical regime. The typical lengths of the production and
regeneration zones is in the range of a few meters. As we can see
from Eqs.~\eqref{axionprob}, \eqref{hpprob} an oscillation length of
a few meters corresponds to particle masses in the range of meV. For
ALPs these experiments are therefore sensitive for masses up to a
few meV and for hidden photons the sensitivity is usually optimal in
this mass range. A variety of these light shining through walls
experiments has been performed or is currently
running~\cite{Ruoso:1992nx,Cameron:1993mr}.
The results can be seen in
Fig.~\ref{lswresult}. The bounds are in the $g\leq {\rm few} \times
10^{-8}\, {\rm GeV}^{-1}$ range for ALPs and the kinetic mixing and
minicharges probed are typically of the order of ${\rm few}\times
10^{-7}$. Although, this is not yet competitive with the
astrophysical and cosmological arguments -- except for the case of hidden
photons in the meV mass range (cf. Fig.~\ref{lswresult} (c)) --
it should be noted that
these bounds are less model-dependent (cf. \cite{Masso:2005ym,Jaeckel:2006xm,Masso:2006gc}).
Moreover, it is worth noting that so far only two of these
experiments (BFRT and ALPS) have employed mirrors to enhance the
transition probability on the production side\footnote{\base Only ALPS
employed a Fabry-Perot cavity whereas BFRT used an optical delay
line. If used only on the production side this does not make a
difference, but for resonant regeneration an optical cavity is
needed.}. Using cavities on both sides~\cite{Hoogeveen:1990vq,Sikivie:2007qm} could lead to significant
improvements in the near future. However, it should be noted that
this presents a technological challenge~\cite{cantatore}
because the cavities have to be tuned to be resonant with each other (and,
of course, a better quality of the cavity reduces the bandwidth
making this even more difficult). Nevertheless, it seems that this
methods paves the way to beat for the first time the astrophysical
bound on the ALP-photon coupling. Moreover, it can even reach
the ALP benchmark point (\ref{g_benchmark_ALP_transp}), $g\sim 10^{-11}$~GeV$^{-1}$,
for small enough ALP mass, therefore possibly testing the ALP
interpretation of the reported hints of cosmic photon regeneration
(cf. Sect.~\ref{Sec:hints_cosmic}).
In order to probe the axion benchmark point (\ref{fa_benchmark_axion_wd}), however, one needs
further ingredients in order to access also higher masses: one possibility
is to exploit alternating magnetic field directions~\cite{VanBibber:1987rq}, another
is to use phase shift plates~\cite{Jaeckel:2007gk}.

Higher masses could also be accessed by exploiting keV photons from X-ray
free-electron lasers or synchrotron radiation sources~\cite{Rabadan:2005dm}.
However, presently the photon
fluxes from these sources appear to be too small to be competitive with
optical photons.

Further improvement in the sensitivity at low masses could also come from
experiments that employ electromagnetic waves in the radiofrequency
range instead of laser light in the visible regime~\cite{Hoogeveen:1992uk}.
The advantage of such ``microwave shining through walls'' is threefold.
First, microwave cavities can be constructed such
that the light is effectively reflected back and forth inside the
cavities up to $10^{11}$ times. This allows for a significant
improvement compared to the currently best optical cavities where
the light can be reflected only up to $10^5$ times. Second, although
still a technological challenge the tuning of the cavities to
resonance is simplified by the fact that extremely frequency-stable
cavities exist in the microwave regime. Finally, using the phase
information of the generator to distinguish between signal and
background one can detect microwaves with intensities less than
$10^{-23}\,{\rm W}$ with commercially available technology. With
currently available technology, sensitivities of the order
$$
g\sim 10^{-10}\,{\rm GeV}^{-1},\quad \chi\sim 10^{-12}
$$
seem achievable. The price to pay is that due to the lower
frequency/energy of the microwaves, the sensitive range is typically
restricted to particle masses less than $0.1\,{\rm meV}$.
Several of these experiments are currently under construction or in
planning stages~\cite{Yale,ADMXcham}.

Finally another variant of the LSW idea is to use static magnetic fields~\cite{Jaeckel:2008sz}.
The setup consists of a highly sensitive magnetometer inside a superconducting shielding.
This is then placed inside a strong (but sub-critical) magnetic field. In ordinary electrodynamics the magnetic field cannot
permeate the superconductor and no field should register on the magnetometer. However, photon -- hidden-sector photon -- photon oscillations would allow to penetrate the superconductor and the magnetic field would ``leak" into the shielded volume and register on the magnetometer.
Compared to the classic LSW setup there are two crucial differences. First, the fields are (nearly) static and the photons involved are virtual.
Second, the magnetometer directly measures the field-strength and not a probability. Correspondingly, the signal is suppressed
only quadratically, $\propto \chi^2$,  instead of quartically.
For hidden photon masses in the range 0.002-200 meV the projected sensitivity for the mixing parameter lies in the
$5\times 10^{-9}$ to $10^{-6}$ range.

\subsubsection{Afterglow Experiments}
For chameleon particles (s. Sect.~\ref{chameleonsect}) and in general particles whose mass increases with the local matter/energy density~\cite{Jaeckel:2006xm}, LSW experiments do not
always work. The reason is that the high (compared to the vacuum) density inside the wall increases the mass of the produced WISP
and therefore creates a high potential barrier on which the particle is reflected.

\begin{figure}[!t]
\includegraphics[width=4.2cm]{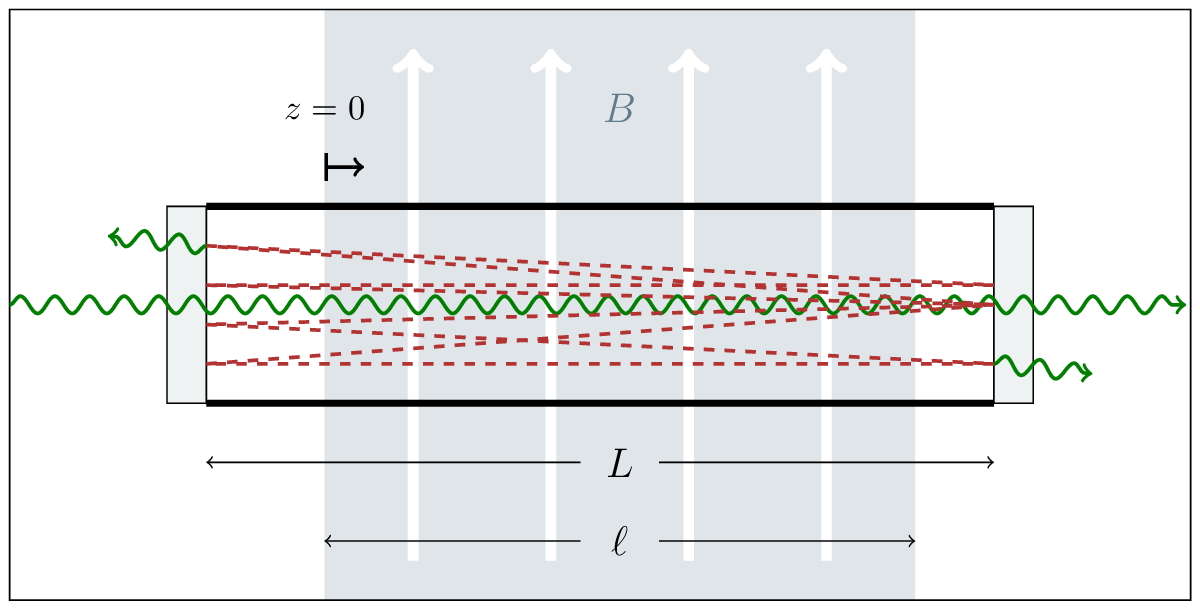}
\hspace*{0.2cm}
\includegraphics[width=4.2cm]{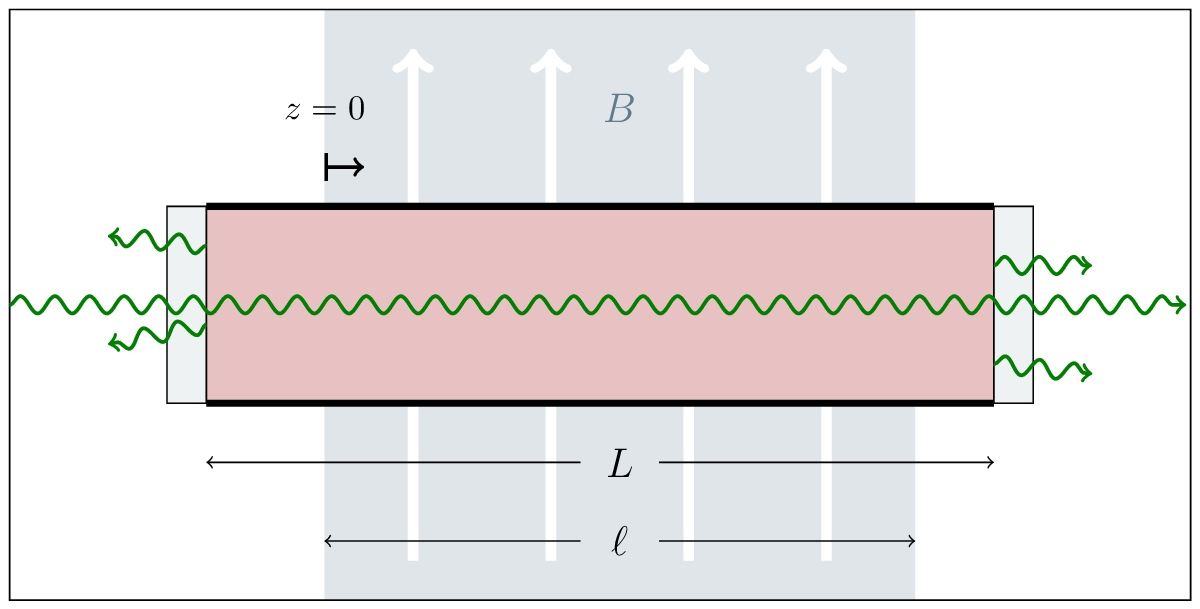}
\hspace*{0.2cm}
\includegraphics[width=4.2cm]{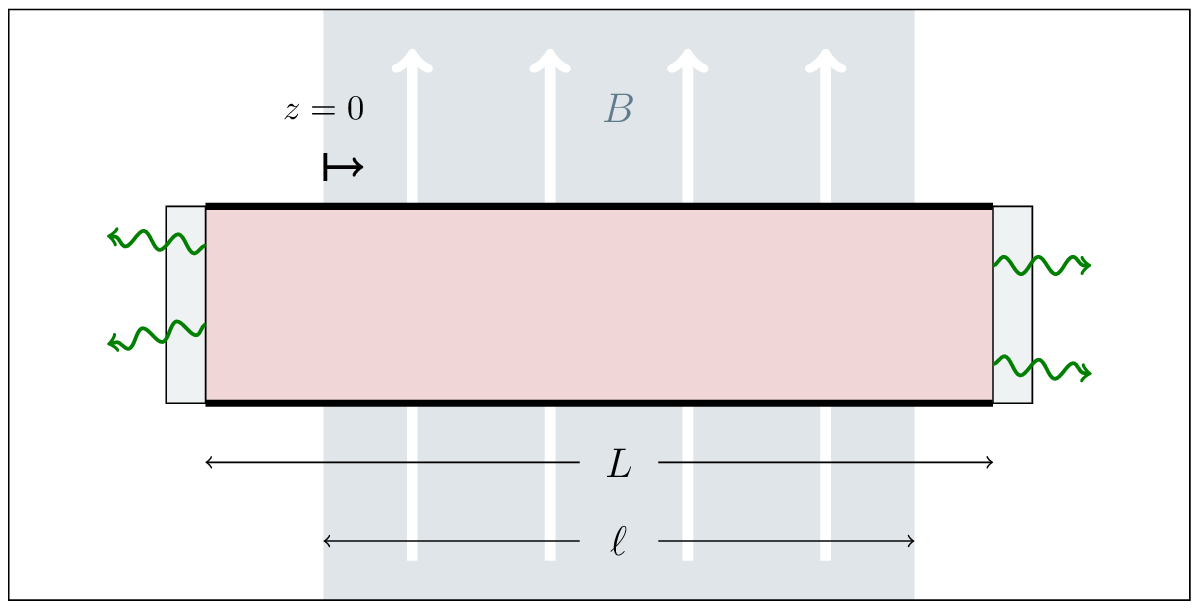}
\caption[]{\base Illustration of an afterglow experiment to search for chameleon particles
(from Ref.~\cite{Ahlers:2007st}).
Filling the vacuum tube by means of a laser beam with chameleons
via photon-chameleon conversion in a magnetic field (left).
An isotropic chameleon gas forms (middle).
Afterglow from chameleon-photon conversion in a magnetic field (right).}
\label{fig:Setup}
\end{figure}

However, this reflection can actually be used in a slightly different type of experiment~\cite{Ahlers:2007st,Gies:2007su}. If the production zone is enclosed by suitably
dense walls on all sides (transparent windows let the photons into the production zone but they are typically too dense to let chameleons escape)
the chameleons produced are trapped inside the production zone and accumulate over time.
After some time the laser is switched off. Now the accumulated chameleons can reconvert into photons which escape through the windows. Hence,
there is an ``afterglow'' after the laser has been switched off. This is shown in Fig.~\ref{fig:Setup}.

Such an experiment has been performed by the GammeV collaboration~\cite{Chou:2008gr} and allows to exclude sub-meV (vacuum mass) chameleons
with couplings in the range $g\sim (10^{-7}-10^{-6})\ {\rm GeV}^{-1}$. Improvements planned by the GammeV collaboration~\cite{gammev}
will further increase the sensitivity. Moreover, by driving the ADMX cavity with an external generator and looking for a ``microwave afterglow'' the ADMX collaboration
has performed a small test experiment for chameleons with masses around $2\,\mu$eV~\cite{ADMXcham}.

\subsubsection{Helioscopes}

As already mentioned in Sect.~\ref{stellar} WISPs can also be
produced inside the sun. This can either occur via the same interactions that lead to the
oscillation phenomena described above or via additional derivative interactions with matter (e.g. for axions).
In any case due to the high number of photons inside the sun and the high total number of interactions,
this typically would lead to the production of a huge number of WISPs.
Helioscopes try to detect these WISPs on earth~\cite{Sikivie:1983ip,vanBibber:1988ge}.

Basically helioscopes employ the same idea as an LSW experiment. However, the production side is replaced by the
WISP production inside the sun. The regeneration side is the same as for a completely laboratory based LSW experiment with the detector pointing towards the
sun. The wall is simply everything in between the solar core and the regeneration side (the rest of the sun, the atmosphere, the walls of the experimental hall etc.).
The enormous total number of interactions inside the sun leads to a large WISP flux even if the coupling is tiny. This
make helioscopes an extremely powerful tool to search for WISPs, however, the same caveats discussed in Sect.~\ref{stellar} apply: if somehow
the production of WISPs inside the sun is suppressed, helioscopes loose their sensitivity.

The typical energy of photons inside the sun is in the keV range, accordingly this is also the typical energy of WISPs produced inside the sun.

Currently two axion helioscopes are running, CAST and SUMICO~\cite{Andriamonje:2007ew,Inoue:2008zp}. The two experiments employ large magnets. Therefore, they are sensitive
to ALPs as well as hidden photons (with and without additional MCPs).
Both experiments have performed searches for
WISPs with keV energies leading to X-ray photons inside the detector. In addition, CAST has also searched for WISPs in the eV regime~\cite{karuza} which
gives weaker but less model dependent constraints on WISPs (cf.~\cite{Jaeckel:2006xm}).

CAST, has recently surpassed the HB constraints for ALPs with a two
photon coupling~\cite{Andriamonje:2007ew} (cf. Fig.~\ref{Fig:axions_astro} (bottom)), and its results have been
used to limit a possible solar $\gamma^\prime$
flux~\cite{Redondo:2008aa,Gninenko:2008pz}.
This is shown in Fig.~\ref{Fig:hp_astro} as part of the purple area. Additional improvements of the CAST apparatus are already in planning~\cite{ferrer}.

Further dedicated helioscopes, an add on to SUMICO~\cite{DedHelio} and a stand alone hidden photon helioscope SHIPS~\cite{SHIPS},
are likely to increase the sensitivity for hidden photons towards smaller masses.

To search for solar axions one could also use the magnetic field of earth for the regeneration and an X-ray satellite
to look through the earth at the solar core~\cite{Davoudiasl:2005nh}.
Alternatively, one could also replace the sun as the source of axions with other cosmic sources~\cite{Fairbairn:2007vj}.

\subsubsection{Direct Axion Dark Matter Searches}
Finally, let us turn to another type of photon regeneration experiment that employs an external source of WISPs: axion dark matter searches.
As the name suggests these experiments are mainly focussed on axions.

As described in Sect.~\ref{axiondm} axions are produced in the early universe and can form all or part of the dark matter.
One can now place the regeneration side of an LSW experiment (with magnet) on earth and wait for dark matter axions to enter the experiment
and be converted into photons. This is the basic idea of an axion dark matter experiment also called axion haloscope~\cite{Sikivie:1983ip}.
In principle this type of experiment is also sensitive to other types of WISPs such as hidden photons. However, due to their different
production mechanisms these typically can only form a small part of the dark matter (which itself puts a constraint on their existence)
which, in addition is typically rather hot, somewhat limiting the sensitivity of haloscope searches for non-axion WISPs. (But a final analysis is still
outstanding.)

The ADMX~\cite{Duffy:2006aa} and CARRACK~\cite{Tada:1999tu} collaborations are currently operating axion haloscopes.
Since the typical masses/energies expected for axion dark matter
lie in the $(1-100)\,\mu$eV range this experiment uses microwave cavities for resonant regeneration (cf. Sect.~\ref{lswexperiments}).
Since dark matter axions are very cold, i.e. their kinetic energy is very small compared to their mass, the
energy of the regenerated photons is basically given by the axion mass. To achieve resonant regeneration the
axion mass therefore has to lie within the bandwidth of the microwave cavities. Since the axion mass is unknown such an experiment scans through a range
of masses by changing the resonance frequency of the cavity.
All in all, as can be seen from Fig.~\ref{Fig:axions_astro}, ADMX achieves enormous sensitivity but currently only for a relatively limited
range of axion masses. Yet, future improvements will significantly increase the scanning speed. With their Phase II upgrade, which
is currently being implemented, ADMX plans to scan the whole range from $10^{-6}$~eV to $10^{-5}$~eV.
It should be noted, however, that regeneration experiments for dark matter axions work only in presence of a background of
axions. Usually, the latter is assumed to be of the size of the local dark matter density and the ADMX bound given in Fig.~\ref{Fig:axions_astro} is obtained under this
assumption.

As an alternative to a microwave cavity experiment one could also
search for dark matter axions using optical cavities filled with laser light.
Photons absorbing dark matter axions would lead to sidebands in the spectrum of the laser light~\cite{Melissinos:2008vn}.

At very small masses (in the finetuning or anthropic region where the initial value of the axion field needs to be small
(cf. Sect.~\ref{axiondm})) the cavities become too large to be practical.
A proposal is to use instead resonant LC-circuits~\cite{Thomas}. This would allow to explore axion dark matter with decay constants up to the GUT scale,
$f_{a}\sim 10^{16}\,{\rm GeV}$.

\subsection{Laser Polarisation Experiments\label{laserpolarisation}}

The experiments described in the previous section search for regenerated photons as a signature of WISPs.
An alternative is to search for the \emph{disappearance} or phase modifications of photons
that would signify the real and virtual production of WISPs~\cite{Maiani:1986md,Gies:2006ca,Ahlers:2007rd}.

This idea is realized in laser polarization experiments.
In these experiments (linearly) polarized laser light is shone through a magnetic field and changes in the polarization are searched for.
The detectable changes are a rotation of the linearly polarized light and also a phase shift between different polarization
directions that modifies linear polarized light into elliptically polarized light.
However, changes in the overall magnitude of the laser field are typically not detectable.

\begin{figure}[t!]
\hspace{-0.7cm}
\subfigure[]{
\scalebox{0.8}[0.8]{
\begin{picture}(240,150)(0,0)
\includegraphics[width=9cm]{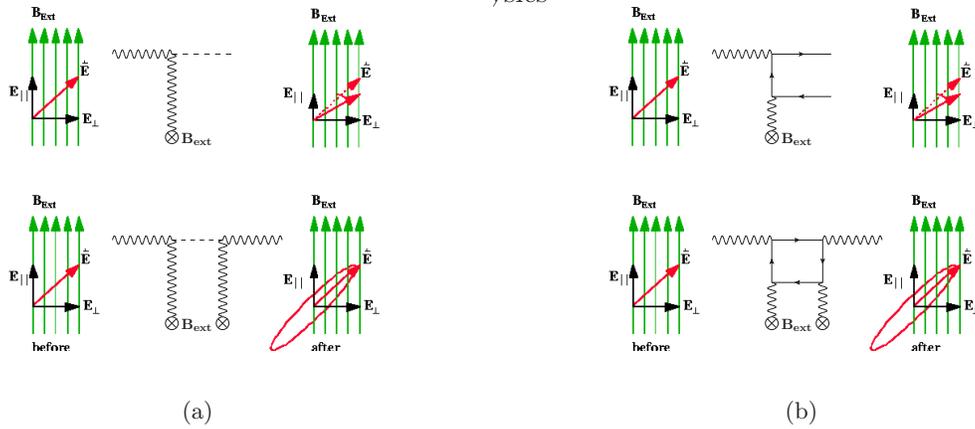}
\scalebox{0.4}[0.4]{
\SetOffset(-460,260)
\Text(100,-10)[c]{\scalebox{1.6}[1.6]{$\mathbf{B}_{\mathbf{ext}}$}}
\Photon(0,90)(70,90){5}{7.5}
\Photon(70,90)(70,00){5}{10.5}
\DashLine(70,90)(140,90){6}
\SetOffset(-460,262)
\CArc(70,-10)(7.5,0,360)
\Line(75,-5)(65,-15)
\Line(65,-5)(75,-15)
\SetOffset(-460,40)
\Text(100,-10)[c]{\scalebox{1.6}[1.6]{$\mathbf{B}_{\mathbf{ext}}$}}
\Photon(0,90)(70,90){5}{7.5}
\Photon(70,90)(70,0){5}{10.5}
\DashLine(70,90)(130,90){6}
\SetOffset(-460,42)
\CArc(70,-10)(7.5,0,360)
\Line(75,-5)(65,-15)
\Line(65,-5)(75,-15)
\SetOffset(-460,40)
\Photon(130,90)(130,0){-5}{10.5}
\SetOffset(-400,42)
\CArc(70,-10)(7.5,0,360)
\Line(75,-5)(65,-15)
\Line(65,-5)(75,-15)
\SetOffset(-460,40)
\Photon(130,90)(200,90){5}{7.5}}
\end{picture}}
\label{alpconversion}}
\hspace{0.8cm}
\subfigure[]{
\scalebox{0.8}[0.8]{\begin{picture}(240,150)(0,0)
\includegraphics[width=9cm]{stream.eps}
\scalebox{0.4}[0.4]{
\SetOffset(-460,260)
\Text(100,-10)[c]{\scalebox{1.6}[1.6]{$\mathbf{B}_{\mathbf{ext}}$}}
\Photon(0,90)(70,90){5}{7.5}
\Photon(70,40)(70,0){5}{4.5}
\ArrowLine(70,90)(140,90)
\ArrowLine(70,40)(70,90)
\ArrowLine(140,40)(70,40)
\SetOffset(-460,262)
\CArc(70,-10)(7.5,0,360)
\Line(75,-5)(65,-15)
\Line(65,-5)(75,-15)
\SetOffset(-460,40)
\Text(100,-10)[c]{\scalebox{1.6}[1.6]{$\mathbf{B}_{\mathbf{ext}}$}}
\Photon(0,90)(70,90){5}{7.5}
\Photon(70,40)(70,0){5}{4.5}
\ArrowLine(70,90)(130,90)
\ArrowLine(70,40)(70,90)
\ArrowLine(130,40)(70,40)
\ArrowLine(130,90)(130,40)
\SetOffset(-460,42)
\CArc(70,-10)(7.5,0,360)
\Line(75,-5)(65,-15)
\Line(65,-5)(75,-15)
\SetOffset(-460,40)
\Photon(130,40)(130,0){-5}{4.5}
\SetOffset(-400,42)
\CArc(70,-10)(7.5,0,360)
\Line(75,-5)(65,-15)
\Line(65,-5)(75,-15)
\SetOffset(-460,40)
\Photon(130,90)(200,90){5}{7.5}}
\end{picture}}
\label{milliconversion}}
\vspace{-0.6cm}
\caption{\base
Rotation (upper half) and ellipticity (lower half) caused by the existence of a light neutral spin-0
boson (left) or a light particle with a small electric charge (right) (figure adapted from \cite{Brandi:2000ty}).}
\label{polarization}
\end{figure}

As an example the relevant processes for ALPs and minicharged particles are depicted in Fig.~\ref{polarization}.

To calculate the relevant changes in phase and amplitude one can use the same equations of motion as in the previous subsection.
Instead of the photon to WISP transition amplitude the relevant quantity is now the photon to photon amplitude and
the corresponding probability $P(\gamma\to\gamma)$.

Note that in order to obtain an observable effect the changes in the amplitude and phase
of the photons must be different for the different polarizations since changes in the overall magnitude of the laser field are not detectable.
Therefore, in presence of Lorentz symmetry these experiments need
a magnetic field that distinguishes a certain axis in order to generate an observable effect.
For the same reason massive hidden photons without any additional matter, which have a magnetic field and polarization independent mass matrix, Eq.~\eqref{hiddenmass},
are not detectable in laser polarization experiments.

In contrast to LSW experiments which have no Standard Model background\footnote{\base Actually this is not quite true. There is a (tiny) background
from the process photon -- graviton -- photon and an additional, wall thickness dependent one, from neutrino tunneling~\cite{Gies:2009wx}.}
there are two processes in QED which provide a background in laser polarization experiments. The first and most prominent is
the QED contribution to the phase shift
. The relevant diagram is the same as in the lower half of Fig.~\ref{milliconversion} just
with ordinary electrons in the loop. Current experiments are only about two orders of magnitude away from the sensitivity required
to measure this effect and the next generation indeed aims to measure it. This limits searches for ALPs via this process
to about $g\gtrsim 10^{-7}\ {\rm GeV}^{-1}$ and for minicharged particles to $\epsilon\gtrsim 10^{-7}$.
The situation is much better for the rotation. Here, the Standard Model background results
from photon splitting, graviton and neutrino production
.
These backgrounds are much much smaller, basically allowing background free discovery potential down to
couplings $g\lesssim 1/M_{P}$ and minicharges $\epsilon \lesssim 10^{-20}$~\cite{Ahlers:2008jt}.
One should note however, that spectacular experimental progress would be needed in order to achieve these sensitivities.

Currently a variety of these experiments have already finished data taking, are running, in construction
or in preparation~\cite{Cameron:1993mr,Zavattini:2005tm}.
Although their basic principle is the same
they differ in that they use pulsed or constant lasers and or magnetic fields.

The polarization experiments described in this section typically only measure the difference in the speed of light between
the two polarization directions. However, using an interferometer where one arm is placed in a magnetic field allows to measure
the change in the speed of light for both polarizations individually. One interesting possibility for this is to equip one
arm of a gravitational wave detector with pulsed magnetic fields~\cite{Dobrich:2009kd}.

In an inhomogeneous magnetic field axions would cause one polarization direction of the laser beam to be split into two parts (similar to a Stern-Gerlach experiment).
Potentially this could also be exploited experimentally to search for axions~\cite{Guendelman:2008jm}.

\subsection{Experiments Using Strong Electromagnetic Fields}
So far we have described the propagation of light using linear equations of motion. All (possibly non-linear) effects of background fields
were included in effective mass matrices that lead to small modifications of the propagation in vacuum.
However, in strong electromagnetic fields non-linear quantum effects can lead to entirely new processes such as, e.g. Schwinger pair production~\cite{Schwinger:1951nm}.
In presence of light particles interacting with the electromagnetic field these effects may set in at much smaller fields than
in ordinary QED where the critical field strength for such effects is set by the electron mass $\sim m^{2}_{e}/e$.
This can be used to search for WISPs. As an example let us consider how minicharged particles can be searched for using Schwinger pair production.

\begin{figure}[t!]
\begin{center}
\includegraphics[angle=0,width=1.0\textwidth]{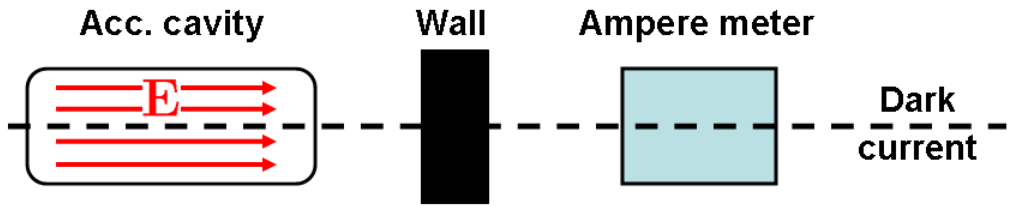}
\end{center}
\vspace{-0.6cm} \caption{
\base
Schematic illustration of an
\emph{accelerator cavity dark current} (AC/DC) experiment
 for searching minicharged particles (from Ref.~\cite{Gies:2006hv}).}
\label{acdc}
\end{figure}

The basic setup \cite{Gies:2006hv} is depicted in Fig.~\ref{acdc}.
In a strong electric field a vacuum pair of charged particles gains energy if the particles are separated by a distance along the lines of the electric field.
If the electric field is strong enough (or the distance large enough) the energy gain can overcome the rest mass, i.e. the virtual particles turn into real particles.
This is the famous Schwinger pair production mechanism~\cite{Schwinger:1951nm}.
After their production the electric field accelerates the particles and antiparticles according to their charge in opposite directions.
This leads to an electric current (dashed line in Fig.~\ref{acdc}).
If the current is made up of minicharged particles the individual particles have very small charges and interact
only very weakly with ordinary matter. Therefore, they can pass even through thick walls nearly unhindered. An electron current, however, would be stopped.
After passage through the wall we can then place an ampere meter to detect the minicharged particle current.

Typical accelerator cavities achieve field strengths of $\gtrsim 25\,{\rm MeV/m}$ and their size is typically of the order of tens of
centimeters.
Precision ampere meters can certainly measure currents as small as $\mu {\rm A}$ and even smaller currents of the order of ${\rm pA}$ seem feasible.
Using the Schwinger pair production
rate we can then estimate the expected sensitivity for such an experiment to be
\begin{equation}
\epsilon_{\rm sensitivity} \sim 10^{-8}\div 10^{-6},\quad{\rm for}\,\,m_{\epsilon}\lesssim {\rm meV}.
\end{equation}
Therefore such an experiment has the potential for significant improvement over the currently best
laboratory
bounds~\cite{Gies:2006ca,Badertscher:2006fm
,Jaeckel:2009dh}, $\epsilon\lesssim{\rm few}\,\, 10^{-7}$ (cf. Fig.~\ref{Fig:mcp_astro}).

\subsection{Fifth Force Experiments\label{Sec:fifth_force}}

Some of the WISP candidates discussed here, such as scalar ALPs or
hidden photons, mediate long range forces between macroscopic
bodies. Therefore, on account of the cumulative effect of a
macroscopic amount of particles, experiments testing the inverse
square law of the Newton and Coulomb forces offer very sensititve
probes of WISPs in a certain mass range~\cite{Adelberger:2009zz}.

Note, that in our discussions of fifth forces we concentrate on
WISPs that interact with photons.
However, searches for
non-Newtonian forces are also sensitive to a whole variety of WISPs
that interact gravitationally or with very weak (effective) Yukawa
interactions. Examples include the dilaton and other moduli,
Kaluza-Klein gravitons but also the chameleons mentioned in
Sect.~\ref{chameleonsect}. For a review
see~\cite{Adelberger:2009zz}.

\begin{figure}[t!]
\begin{center}
\scalebox{0.5}[0.5]{
  \begin{picture}(258,162) (111,-95)
    \SetWidth{1.0}
    \SetColor{Black}
    \Line[arrow,arrowpos=0.5,arrowlength=5,arrowwidth=2,arrowinset=0.2](112,66)(160,18)
    \Line[arrow,arrowpos=0.5,arrowlength=5,arrowwidth=2,arrowinset=0.2](160,18)(160,-46)
    \Line[arrow,arrowpos=0.5,arrowlength=5,arrowwidth=2,arrowinset=0.2](160,-46)(112,-94)
    \Photon(160,-46)(208,-14){7.5}{2}
    \Photon(160,18)(208,-14){-7.5}{2}
    \Line[dash,dashsize=2](208,-14)(272,-14)
    \Photon(272,-14)(320,-46){-7.5}{2}
    \Photon(272,-14)(320,18){7.5}{2}
    \Line[arrow,arrowpos=0.5,arrowlength=5,arrowwidth=2,arrowinset=0.2](368,-94)(320,-46)
    \Line[arrow,arrowpos=0.5,arrowlength=5,arrowwidth=2,arrowinset=0.2](320,-46)(320,18)
    \Line[arrow,arrowpos=0.5,arrowlength=5,arrowwidth=2,arrowinset=0.2](320,18)(368,66)
  \end{picture}
}
\end{center}
\caption{\base
Exchange of a scalar ALP, coupled two photons via Eq.~(\ref{Lone}),
between two protons, giving rise to a Yukawa-type non-Newtonian force between two neutral test bodies
(from Ref.~\cite{Adelberger:2006dh}).
\label{fig:non_newtonian}}
\end{figure}
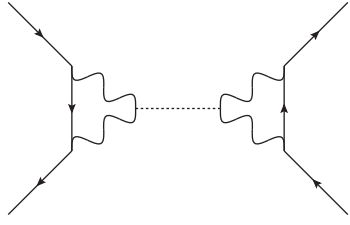

\subsubsection{Bounds on ALPs from Searches for Non-Newtonian Forces}

Due to its pseudoscalar (pseudo-)Goldstone boson nature, the exchange of a very light axion
dominantly leads to a spin-dependent force which can not be probed by the most sensitive
experiments exploiting unpolarized bodies. A spin-independent force is generated by the
exchange of two axions\footnote{\base Single pseudoscalar exchange leads to long range interactions between
unpolarized test bodies only in presence of CP-violation. For axions
the resulting interaction potential is $\sim \bar{\theta}^2$~\cite{Moody:1984ba,Adelberger:2009zz}.
Although $\bar{\theta}$ is expected to be non-vanishing due to the CP-violation in the weak interactions
its value is very small (the electric dipole moment of the neutron after all tells us $\bar{\theta}<3\times10^{-10}$).},
which leads to a power-law correction to the spin-independent
long-range force between neutral test bodies \cite{Ferrer:1998ue}. The bounds inferred from
corresponding torsion-balance type experiments on $f_a$ are, however, not competitive with
astrophysical bounds~\cite{Adelberger:2006dh}.

\begin{figure}[t!]
\centerline{\includegraphics[width=.8\textwidth]{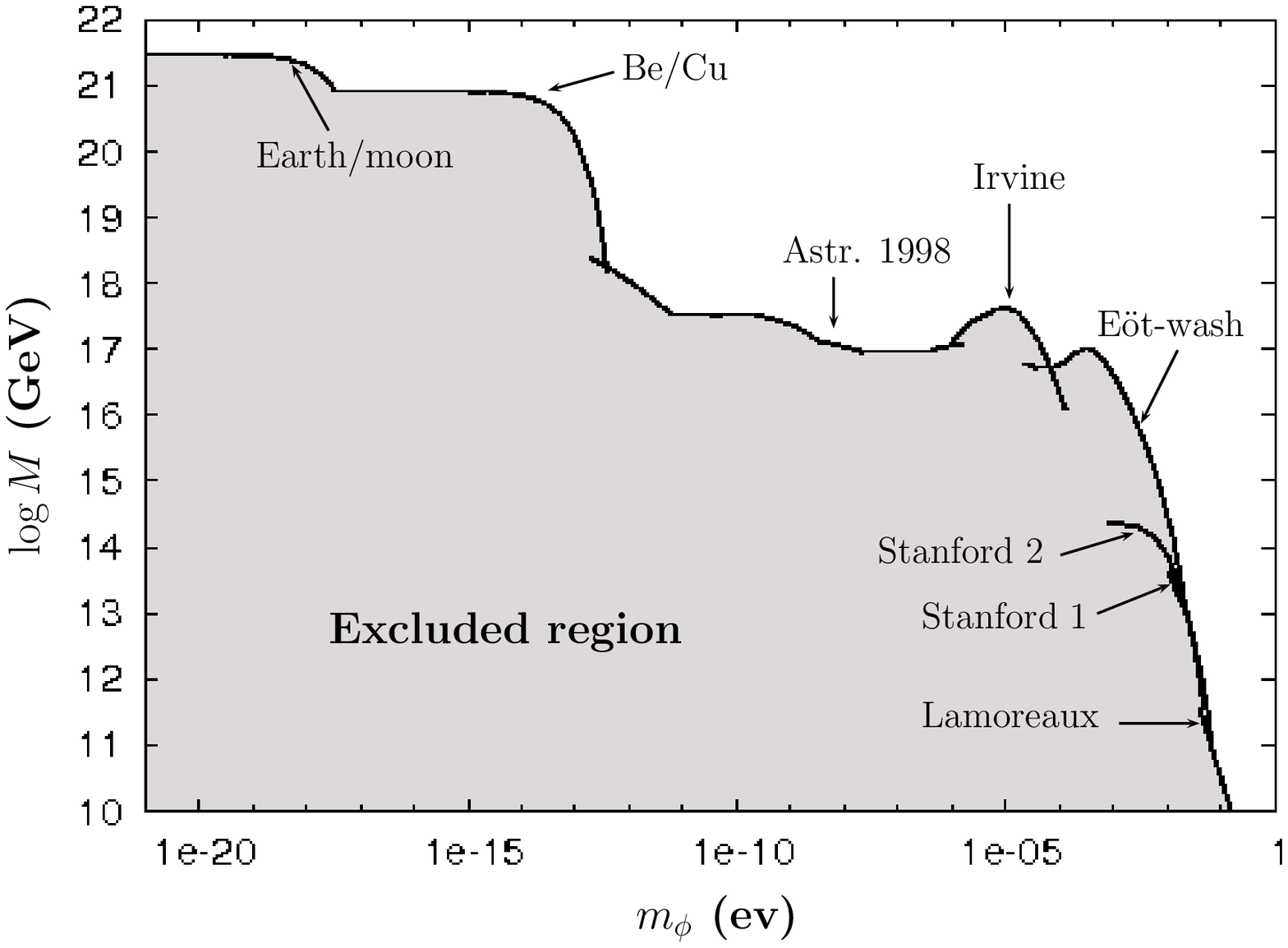}}
\caption{\base
Constraints on the two photon coupling $M=1/g$ of a scalar ALP vs.
its mass $m_\phi$ from searches for non-Newtonian forces exploiting torsion balance
techniques (Be/Cu~\cite{Su:1994gu}, Irvine~\cite{Spero:1980zz},
E\"ot-wash~\cite{Kapner:2006si,Adelberger:2006dh},
Stanford~\cite{Chiaverini:2002cb}) or Casimir force searches
(Lamoreaux~\cite{Lamoreaux:1996wh}). Also shown are constraints from
astrophysical observations (Earth/moon~\cite{Dickey:1994}, Astr. 1998~\cite{Onofrio:2006mq}).
Compilation from Ref.~\cite{Dupays:2006dp}.
\label{fig:non_newtonian_M}}
\end{figure}

This is different for scalar ALPs, whose coupling to two photons occurs via
the effective Lagrangian
\begin{equation}
{\cal L}_1= {g \over 4}\, \phi F^{\mu\nu} F_{\mu\nu}  .
\label{Lone}
\end{equation}
This coupling leads, via the radiative corrections shown in Fig.~\ref{fig:non_newtonian},
to a spin-independent non-Newtonian force between test bodies of the Yukawa-type,
\begin{equation}
\propto (gm_p)^2\exp(-m_\phi r).
\end{equation}
From the non-observation of such a force in sensitive
torsion-balance searches for Yukawa violations of the gravitational
inverse-square law one may put a very stringent
limit~\cite{Dupays:2006dp,Adelberger:2006dh},
\begin{equation}
g\lesssim 4\times 10^{-17}\ {\rm GeV}^{-1},
\end{equation}
in the meV mass range and even stronger constraints for smaller masses
(see also Fig.~\ref{fig:non_newtonian_M}).

In order to test for the pseudoscalar axions one can also search for monopole-(spin)dipole ($\sim \bar{\theta}^{1}$) and dipole-dipole
interactions ($\sim \bar{\theta}^{0}$)~\cite{Moody:1984ba,Adelberger:2009zz} (see \cite{Vasilakis:2008yn} for some very recent new data),
where $\bar{\theta}$ is the remaining CP violating angle due to the CP violation in the electroweak sector.
However, for monopole-dipole searches again the smallness of $\bar{\theta}$ limits the sensitivity and dipole-dipole searches are experimentally rather challenging.

\subsubsection{Bounds on WISPs from Searches for Non-Coulomb Forces}\label{fifthforces}

Hidden photons~\cite{Okun:1982xi,Popov:1999} and minicharged particles~\cite{Jaeckel:2009dh} also leave detectable imprints as modifications of Coulomb's law.

For massive hidden photons the non-diagonal mass term in the equations of motion leads to a modification
of the potential between two charges,
\begin{equation}
V(r) = \frac{\alpha}{r}\left( 1 + \chi^2 e^{-m_{\gamma^\prime}r}\right).
\end{equation}
Minicharged particles modify the potential at the one-loop level by the Uehling contribution. At large
distances the deviation is exponential as well,
\begin{equation}
\label{Uehlingapprox}
V(r)\approx  \frac{\alpha}{r} \left[1+ \frac{\alpha\epsilon^2}{4\sqrt{\pi}}\frac{\exp(-2mr)}{(mr)^{\frac{3}{2}}}\right]
,\quad\quad\quad\quad\quad\quad{\rm for}\quad mr\gg 1.
\end{equation}

The inverse square-law has been tested in the laboratory by
Cavendish-type experiments, checking for the absence of an electric field inside a
charged conducting sphere~\cite{Williams:1971ms}. These experiments,
originally performed to set limits on the photon mass~\cite{Goldhaber:1971mr} --
which would also give an exponential
deviation from the Coulomb law -- and to constrain also the range and strength of
an electrical fifth force~\cite{Bartlett:1988yy}, give also a strong constraint on
hidden photons, notably in the $\mu$eV range (cf. Fig.~\ref{Fig:hp_astro},
labeled ``Coulomb"). Similarly, they give the best laboratory constraints on minicharged particles in the sub-$\mu$eV range.

This demonstrates that searches for deviations from Coulomb's law are a powerful tool to search for WISPs, giving strong motivation to
improve upon the by now nearly 40 year old experiments.

Static magnetic fields are also modified in presence of hidden photons. In particular, the large scale magnetic fields
of planets can be used to test for hidden photons with very small masses as can be seen from the constraints
``Earth" or ``Jupiter" in Fig.~\ref{Fig:hp_astro} (cf.~\cite{Goldhaber:1968mt,Goldhaber:1971mr}).

\section{Summary and Outlook}\label{conclusions}
For particle physicists the term ``low energy physics'' is usually used for physics below the current high energy frontier, today roughly
the electroweak scale $\sim 100\,{\rm GeV}$. Currently, three areas of ``low energy physics'' are proving to
be fruitful grounds to explore fundamental physics.
First there are experiments just below the electroweak scale, in the regime of roughly $1\,{\rm GeV}$-$100\,{\rm GeV}$. This energy range is particularly interesting for high precision flavor physics
but it could also provide insight into so-called
``Dark Forces'' that have recently attracted a lot of attention in order to resolve puzzling astrophysical observations
(cf. Sect.~\ref{Sec:dm_hidden}).
The second area of interest is to use low energy experiments and observations to test fundamental symmetries, such as e.g. Lorentz symmetry, to an incredibly
high precision (for a review see~\cite{Kostelecky:2007qf}).
Finally, there is the possibility that there exist new particles with very small masses (possibly even sub-eV) but also very weak interactions
with the known Standard Model particles. This is the case discussed in this review.

Light weakly coupled particles with masses below an eV, so-called weakly interacting sub-eV particles (WISPs), are strongly motivated both
from top down as well as from bottom-up considerations. A classic example for a WISP is the axion
(cf. Sect.~\ref{Sec:QCD_axion}).
From a bottom-up point of view the axion is predicted as a consequence of the Peccei-Quinn
solution of the strong CP problem. A new
global symmetry is introduced which is spontaneously broken at a very high energy scale $\sim f_{a}$. As a consequence, the axion as a pseudo-Nambu-Goldstone boson of this symmetry
has both small mass, $m_{a}\sim {\rm meV}(10^{10}\, {\rm GeV}/f_{a})$, and coupling to two photons, $g\sim 10^{-13}\ {\rm GeV}^{-1}(10^{10}\, {\rm GeV}/f_{a})$.
Taking, on the other hand, a top-down approach, in string theory particles with couplings similar to axions, axion-like particles,
seem to be a natural consequence of the compactification of extra dimensions (cf. Sect.~\ref{Sec:axions_string}).
Again, their small couplings arise from the large energy scales involved, the string scale (possibly also the sizes of the extra dimensions).
Searching for such particles is thereby a way to probe very high energy scales far beyond the energy scale in colliders such as LHC.

Axions are by far not the only possible WISP candidates. String models often contain also additional U(1) gauge groups and matter charged under
them (cf. Sect.~\ref{Sec:hidden_string}).
The corresponding extra ``hidden'' photons can mix kinetically with the ordinary electromagnetic photon. Small mixing angles and corresponding small interactions with ordinary matter
arise from high energy scales involved and from the fact that the additional gauge factors are ``far away" in the extra dimensions. Roughly speaking
probing small couplings allows us look beyond our immediate neighborhood and probe the global structure of the compactification.
For extra photons (and matter charged under them) small masses are possible but not necessary.

Astrophysics and cosmology already provide a powerful tool to constrain WISPs and in many regions of parameter space this
sets the standard against which laboratory tests have to measure up (cf. Sect.~\ref{astro}).
The rapidly increasing amount of available data will hopefully further
improve the situation (see also Ref.~\cite{Chelouche:2008ta}).

Beyond this some astrophysical and cosmological observations even provide suggestive hints for the existence of WISPs
(cf. Sect.~\ref{hints}).
Importantly, both axions as well as hidden photons (or particles related to them) may contribute all or part of the dark matter.
In addition to the above there is a number of astrophysical puzzles that can be solved by the presence of WISPs such as, e.g., the
possible observations of high and very high energy cosmic photon regeneration,
the observed alignment of the polarization vectors of very distant quasars,
the energy loss of white dwarfs, and a slight excess in the effective number of neutrino species in the CMB measurements.
These interpretations can and should be tested in the laboratory.

One of the most important features of the WISPs described above is that they have (very weak) interactions with photons.
In combination with their small mass this allows to search for them in photon regeneration experiments,
in laser polarisation experiments, in experiments exploiting strong electromagnetic fields, and also in
fifth force experiments.

Currently, considerable activity takes place in the field of laser light shining through a wall experiments
(cf. Sect.~\ref{lswexperiments}),
which is now entering a new generation.
Important advances in laser technology appear to pave the way to beat the sensitivity of current WISP helioscopes
and to probe the above mentioned explanations of astrophysical puzzles in terms of photon $\leftrightarrow$ WISP oscillations.
Pioneering experiments exploiting instead high-quality microwave cavities for the generation and regeneration of
WISPs are in the commissioning phase. The microwave cavity search for dark matter axions probes a complementary
region in parameter space compared to the other photon regeneration experiments and should provide a definitive
answer whether axions are the dominant part of cold dark matter within the current decade.

Laser polarisation experiments (cf. Sect.~\ref{laserpolarisation}) will continue their quest to detect the QED birefringence. At the same time they
will also considerably improve the bounds they place on WISPs. They will also further help to develop the optical
techniques, such as cavities that are used in light shining through a wall experiments. Moreover, experiments analyzing
the spectrum of laser light in an interferometer also seem to be a promising tool to search for dark matter axions.

Fifth force experiments (cf. Sect.~\ref{Sec:fifth_force}), in particular searches for deviations from the gravitational inverse square law are already
an established tool to test fundamental physics which will continue to improve
significantly, thereby providing new bounds on a variety of WISPs,
in particular light scalar fields. Other WISPs such as hidden photons and minicharged partcles can be probed with high sensitivities
by testing for deviations from Coulomb's law. Although tests of Coulomb's law are already at an impressive precision the currently
best available bounds in the length range of meters are nearly 40 years old. It stands to hope that with current technology considerable improvements
are possible.

All in all using low energy experiments with photons to search for WISPs may give important information about fundamental
particle physics complementary to the one obtainable at high energy colliders.
Already today these experiments provide very strong bounds on light weakly interacting particles.
But even more excitingly the next few years are likely to bring considerable advances and huge discovery potential for new physics.

\section{Acknowledgements}

The authors would like to thank Markus Ahlers, Holger Gies, Mark Goodsell, Axel Lindner, Alessandro Mirrizzi, Javier Redondo and Christoph Weniger
for interesting discussions, helpful suggestions and joyful collaboration on the subjects discussed in this review.




\begin{thebibliography}{99}
\base

\bibitem{Amsler:2008zzb}
  Amsler C, et al.  [Particle Data Group].
  {\it Phys.\ Lett. B} 667:1 (2008)

\bibitem{Peccei:1977hh}
Peccei RD, Quinn HR.
  {\it Phys.\ Rev.\ Lett.}  38:1440 (1977)

\bibitem{Weinberg:1977ma}
Weinberg S.
  {\it Phys.\ Rev.\ Lett.}  40:223 (1978)\\
%
  Wilczek F.
  {\it Phys.\ Rev.\ Lett.}  40:279 (1978)

\bibitem{Kim:1979if}
  Kim JE.
  {\it Phys.\ Rev.\ Lett.} 43:103 (1979)\\
%
  Dine M, Fischler W, Srednicki M.
  {\it Phys.\ Lett.  B} 104:199 (1981)\\
  Shifman MA, Vainshtein AI, Zakharov VI.
  {\it Nucl.\ Phys.\  B} 166:493 (1980)\\
  Zhitnitsky AR.
  {\it Sov.\ J.\ Nucl.\ Phys.\ } 31:260 (1980) 260
  [{\it Yad.\ Fiz.\ } 31:497 (1980)]

\bibitem{Bardeen:1977bd}
  Bardeen WA, Tye SH.
  {\it Phys.\ Lett.\  B} 74:229 (1978)\\
%
  Kaplan DB.
  {\it Nucl.\ Phys.\  B} 260:215 (1985)\\
%
  Srednicki M.
  {\it Nucl.\ Phys.\  B} 260:689 (1985)

\bibitem{Witten:1984dg}
  Witten E.
  {\it Phys.\ Lett.\  B} 149:351 (1984)

\bibitem{Conlon:2006tq}
  Conlon JP.
  {\it JHEP} 0605:078 (2006)
  [arXiv:hep-th/0602233]

\bibitem{Svrcek:2006yi}
  Svrcek P, Witten E.
  {\it JHEP} 0606:051 (2006)
  [arXiv:hep-th/0605206]

\bibitem{Arvanitaki:2009fg}
  Arvanitaki A, et al.
  arXiv:0905.4720 [hep-th]

\bibitem{Choi:1985je}
  Choi K,  Kim JE.
  {\it Phys.\ Lett.\  B} 154:393 (1985);
%
  {\it Phys.\ Lett.\  B} 165:71 (1985)

\bibitem{Wetterich:1987fm}
  Wetterich C.
  {\it Nucl.\ Phys.\  B} 302:668 (1988)\\
  Ratra B, Peebles PJE.
  {\it Phys.\ Rev.\  D} 37:3406 (1988)\\
  Caldwell RR, Dave R, Steinhardt PJ.
  {\it Phys.\ Rev.\ Lett.\ } 80:1582 (1998)
  [arXiv:astro-ph/9708069]

\bibitem{Khoury:2003aq}
  Khoury J, Weltman A.
  {\it Phys.\ Rev.\ Lett.\ } 93:171104 (2004)
  [arXiv:astro-ph/0309300]\\
  Khoury J, Weltman A.
  {\it Phys.\ Rev.\  D} 69:044026 (2004)
  [arXiv:astro-ph/0309411]\\
  Brax P, et al.
  {\it Phys.\ Rev.\  D} 70:123518 (2004)
  [arXiv:astro-ph/0408415]

\bibitem{Candelas:1985en}
  Candelas P, Horowitz GT, Strominger A, Witten E.
  {\it Nucl.\ Phys.\  B} 258:46 (1985)\\
  %
  Witten E.
  {\it Nucl.\ Phys.\  B} 268:79 (1986)



\bibitem{Lebedev:2008un}
  Lebedev O, et al.
  {\it Phys.\ Lett.\  B} 668:331 (2008)
  [arXiv:0807.4384 [hep-th]].

\bibitem{Lebedev:2009ag}
  Lebedev O, Ramos-Sanchez S.
  arXiv:0912.0477 [hep-ph].

\bibitem{Holdom:1985ag}
  Holdom B.
  {\it Phys.\ Lett.\  B} 166:196 (1986)

\bibitem{Dienes:1996zr}
  Dienes KR, Kolda CF, March-Russell J,
  {\it Nucl.\ Phys.\  B} 492:104 (1997)

\bibitem{Lust:2003ky}
  Lust D, Stieberger S.
  {\it Fortsch.\ Phys.\ } 55:427 (2007)
  [arXiv:hep-th/0302221]
%
  Abel SA, Schofield BW.
  {\it Nucl.\ Phys.\  B} 685:150 (2004)
  [arXiv:hep-th/0311051]\\
%
  Berg M, Haack M, Kors B.
  {\it Phys.\ Rev.\  D} 71:026005 (2005)
  [arXiv:hep-th/0404087]\\
  Abel SA, Jaeckel J, Khoze VV, Ringwald, A.
  {\it Phys.\ Lett.\  B} 666:66 (2008)
  [arXiv:hep-ph/0608248]

\bibitem{Abel:2008ai}
  Abel SA, et al.
  {\it JHEP} 0807:124 (2008)
  [arXiv:0803.1449 [hep-ph]]

\bibitem{Goodsell:2009xc}
  Goodsell M, Jaeckel J, Redondo J, Ringwald, A.
  {\it JHEP} 0911:027 (2009)
  [arXiv:0909.0515 [hep-ph]]

\bibitem{Ahlers:2008qc}
  Ahlers M, Jaeckel J, Redondo J, Ringwald A.
  {\it Phys.\ Rev.\  D} 78:075005 (2008)
  [arXiv:0807.4143 [hep-ph]].

\bibitem{Bruemmer:2009ky}
  Brummer F, Jaeckel J, Khoze VV.
  {\it JHEP} 0906:037 (2009)
  [arXiv:0905.0633 [hep-ph]]

\bibitem{Raffelt:1996wa}
Raffelt GG. {\em Stars as laboratories for fundamental physics: The
  astrophysics of neutrinos, axions, and other weakly interacting particles}.
\newblock The University of Chicago Press, 1996.
\newblock 664 p.

\bibitem{Raffelt:2006cw}
Raffelt GG.
{\it Lect.  Notes Phys.} 741:51 (2008)

\bibitem{Raffelt:1985nj}
Raffelt GG.
{\it Phys. Lett. B} 166:402 (1986)

\bibitem{Isern:2008nt}
Isern J, Garcia-Berro E, Torres S, Catalan S.
arXiv:0806.2807 [astro-ph]

\bibitem{Andriamonje:2007ew}
Andriamonje S, et al.  [CAST Collaboration].
{\it JCAP} 0704:010 (2007)\\
  Arik E, et al.  [CAST Collaboration].
  {\it JCAP} 0902:008 (2009)
  [arXiv:0810.4482 [hep-ex]]

\bibitem{Schlattl:1998fz}
Schlattl H, Weiss A, Raffelt GG.
{\it Astropart. Phys.} 10:353 (1999)

\bibitem{Inoue:2008zp}
  Inoue Y, et al.
  {\it Phys.\ Lett.\  B} 668:93 (2008)
  [arXiv:0806.2230 [astro-ph]]

\bibitem{Redondo:2008en}
  Redondo J.
  arXiv:0810.3200 [hep-ph]

\bibitem{Raffelt:1985nk}
Raffelt GG.
{\it Phys. Rev. D} 33:897 (1986)

\bibitem{Raffelt:1987yu}
Raffelt GG, Dearborn DSP.
{\it Phys. Rev. D} 36:2211 (1987)

\bibitem{Brockway:1996yr}
  Brockway JW, Carlson ED, Raffelt GG.
  {\it Phys.\ Lett.\  B} 383:439 (1996)
  [arXiv:astro-ph/9605197]\\
  Grifols JA, Masso E, Toldra R.
  {\it Phys.\ Rev.\ Lett.\ } 77:2372 (1996)
  [arXiv:astro-ph/9606028]

\bibitem{Gondolo:2008dd}
Gondolo P, Raffelt GG.
arXiv:0807.2926 [astro-ph]

\bibitem{Redondo:priv}
  Redondo J.
  private communication

\bibitem{Redondo:2008aa}
  Redondo J.
  {\it JCAP} 0807:008 (2008)
  [arXiv:0801.1527 [hep-ph]]

\bibitem{Redondo:2008ec}
  Redondo J, Postma M.
  {\it JCAP} 0902:005 (2009)
  [arXiv:0811.0326 [hep-ph]]

\bibitem{Jaeckel:2006xm}
  Jaeckel J, et al.
  arXiv:hep-ph/0605313;
  {\it Phys.\ Rev.\  D} 75:013004 (2007)
  [arXiv:hep-ph/0610203]

\bibitem{Masso:2005ym}
  Masso E, Redondo J.
  {\it JCAP} 0509:015 (2005)
  [arXiv:hep-ph/0504202]\\
  Mohapatra RN, Nasri S.
  {\it Phys.\ Rev.\ Lett.\ } 98:050402 (2007)
  [arXiv:hep-ph/0610068]\\
  Brax P, van de Bruck C, Davis AC.
  {\it Phys.\ Rev.\ Lett.\ } 99:121103 (2007)
  [arXiv:hep-ph/0703243]\\
  Jain P, Mandal S.
  {\it Int.\ J.\ Mod.\ Phys.\  D} 15:2095 (2006)
  [arXiv:astro-ph/0512155]\\
  Jain P, Stokes S.
  arXiv:hep-ph/0611006\\
  Kim JE.
  {\it Phys.\ Rev.\  D} 76:051701 (2007)
  [arXiv:0704.3310 [hep-ph]]

\bibitem{Masso:2006gc}
  Masso E, Redondo J.
  {\it Phys.\ Rev.\ Lett.\ } 97:151802 (2006)
  [arXiv:hep-ph/0606163]\\
  Redondo J.
  arXiv:0807.4329 [hep-ph]


\bibitem{Iocco:2008va}
  Iocco F, Mangano G, Miele G, Pisanti O, Serpico PD.
  {\it Phys.\ Rept.\ } 472:1 (2009)
  [arXiv:0809.0631 [astro-ph]]

\bibitem{Simha:2008zj}
  Simha V, Steigman G.
  {\it JCAP} 0806:016 (2008)
  [arXiv:0803.3465 [astro-ph]]

\bibitem{Davidson:2000hf}
Davidson S, Hannestad S, Raffelt G.
{\it JHEP} 05:003 (2000)
[hep-ph/0001179]\\
  Berezhiani Z, Lepidi A.
  {\it Phys.\ Lett.\  B} 681:276 (2009)
  [arXiv:0810.1317 [hep-ph]]

\bibitem{Fixsen:1996nj}
Fixsen DJ, et al.
  {\it Astrophys. J.} 473:576 (1996)
arXiv:astro-ph/9605054

\bibitem{Melchiorri:2007sq}
  Melchiorri A, Polosa A, Strumia A.
  {\it Phys.\ Lett.\  B} 650:416 (2007)
  [arXiv:hep-ph/0703144]

\bibitem{Jaeckel:2008fi}
  Jaeckel J, Redondo J, Ringwald A.
  {\it Phys.\ Rev.\ Lett.\ }  101:131801 (2008)
  [arXiv:0804.4157 [astro-ph]]

\bibitem{Mirizzi:2009iz}
  Mirizzi A, Redondo J, Sigl G.
  {\it JCAP} 0903:026 (2009) 026
  [arXiv:0901.0014 [hep-ph]].

\bibitem{Mirizzi:2009nq}
  Mirizzi A, Redondo J, Sigl G.
  {\it JCAP} 0908:001 (2009)
  [arXiv:0905.4865 [hep-ph]]

\bibitem{Lawrence}
  Lawrence CR, Herbig T, Readhead ACS, Gulkis S.
  {\it The Astrophysical Journal Letters} 449:L5-L8 (1995)\\
  Bennett C, et al. [WMAP Collaboration].
  {\it Astrophys.\ J.\ Suppl.\ } 148:97 (2003)
  [arXiv:astro-ph/0302208]\\
  de Petris M, et al.
  {\it Astrophys.\ J.\ } 574:L119 (2002)
  [arXiv:astro-ph/0203303]

\bibitem{Burrage:2009yz}
  Burrage C, Jaeckel J, Redondo J, Ringwald A.
  {\it JCAP} 0911:002 (2009)
  [arXiv:0909.0649 [astro-ph.CO]]

\bibitem{Ahlers:2009kh}
  Ahlers M.
  {\it Phys.\ Rev.\  D} 80:023513 (2009)
  [arXiv:0904.0998 [hep-ph]]


\bibitem{Ichikawa:2008pz}
  Ichikawa K, Sekiguchi T, Takahashi T.
  {\it Phys.\ Rev.\  D} 78:083526 (2008)
  [arXiv:0803.0889 [astro-ph]]

\bibitem{cosmoaxion}
  Hannestad S, Mirizzi A, Raffelt G.
  {\it JCAP} 0507:002 (2005)
  [arXiv:hep-ph/0504059]\\
  Melchiorri A, Mena O, Slosar A.
  {\it Phys.\ Rev.\  D} 76:041303 (2007)
  [arXiv:0705.2695 [astro-ph]]\\
  Hannestad S, Mirizzi A, Raffelt GG, Wong YYY.
  {\it JCAP} 0708:015 (2007)
  [arXiv:0706.4198 [astro-ph]]\\
  Hannestad S., et al.
  arXiv:0910.5706 [hep-ph].

\bibitem{Seljak:2006bg}
  Seljak U, Slosar A, McDonald P.
  {\it JCAP} 0610:014 (2006)
  [arXiv:astro-ph/0604335];
  Mangano G, et al.
  {\it JCAP} 0703:006 (2007)
  [arXiv:astro-ph/0612150]

\bibitem{Hamann:2007pi}
  Hamann J, Hannestad S, Raffelt GG, Wong YYY.
  {\it JCAP} 0708:021 (2007)
  [arXiv:0705.0440 [astro-ph]]

\bibitem{Sikivie:2006ni}
  Sikivie P.
  {\it Lect.\ Notes Phys.\ } 741:19 (2008)
  [arXiv:astro-ph/0610440].

\bibitem{Preskill:1982cy}
  Preskill J, Wise MB, Wilczek F.
  {\it Phys.\ Lett.\  B} 120:127 (1983)\\
%
  Abbott LF, Sikivie P.
  {\it Phys.\ Lett.\  B} 120:133 (1983)\\
%
  Dine M, Fischler W.
  {\it Phys.\ Lett.\  B} 120:137 (1983)

\bibitem{Hertzberg:2008wr}
  Hertzberg MP, Tegmark M, Wilczek F.
  {\it Phys.\ Rev.\  D} 78:083507 (2008)
  [arXiv:0807.1726 [astro-ph]].

\bibitem{Tegmark:2003ve}
  Tegmark M, de Oliveira-Costa A, Hamilton A.
  {\it Phys.\ Rev.\  D} 68:123523 (2003)
  [arXiv:astro-ph/0302496]\\
  de Oliveira-Costa A, Tegmark M, Zaldarriaga M, Hamilton A.
  {\it Phys.\ Rev.\  D} 69:063516 (2004)
  [arXiv:astro-ph/0307282]\\
  Copi CJ, Huterer D, Schwarz DJ, Starkman GD.
  {\it Mon.\ Not.\ Roy.\ Astron.\ Soc.\ } 367:79 (2006)
  [arXiv:astro-ph/0508047]

\bibitem{Kinney:1999rk}
  Kinney WH, Sikivie P.
  {\it Phys.\ Rev.\  D} 61:087305 (2000)
  [arXiv:astro-ph/9906049]\\
  Sikivie P.
  {\it Phys.\ Lett.\  B} 567:1 (2003)
  [arXiv:astro-ph/0109296]

\bibitem{Sikivie:2009fv}
  Sikivie P.
  arXiv:0909.0949 [hep-ph]

\bibitem{Hoffmann:1987et}
  Hoffmann S
  {\it Phys.\ Lett.\  B} 193:117 (1987)

\bibitem{Aharonian:2005gh}
  Aharonian F, et al.   [H.E.S.S. Collaboration].
  {\it Nature} 440:1018 (2006)
  [arXiv:astro-ph/0508073]

\bibitem{Mazin:2007pn}
  Mazin D, Raue M.
  {\it Astron.\ Astrophys.\ } 471:439 (2007)
  [arXiv:astro-ph/0701694]

\bibitem{Hochmuth:2007hk}
  Hochmuth KA, Sigl G.
  {\it Phys.\ Rev.\  D} 76:123011 (2007)
  [arXiv:0708.1144 [astro-ph]]\\
  Hooper D, Serpico PD.
  {\it Phys.\ Rev.\ Lett.\ } 99:231102 (2007)
  [arXiv:0706.3203 [hep-ph]]

\bibitem{DeAngelis:2007dy}
  De Angelis A, Mansutti O, Roncadelli M.
  {\it Phys.\ Rev.\  D} 76:121301 (2007)
  [arXiv:0707.4312 [astro-ph]]

\bibitem{Mirizzi:2009aj}
  Mirizzi A, Montanino D.
  {\it JCAP} 0912:004 (2009)
  [arXiv:0911.0015 [astro-ph.HE]]

\bibitem{Burrage:2009mj}
  Burrage C, Davis AC, Shaw DJ.
  {\it Phys.\ Rev.\ Lett.\ } 102:201101 (2009)
  [arXiv:0902.2320 [astro-ph.CO]]

\bibitem{Fairbairn:2009zi}
  Fairbairn M, Rashba T, Troitsky S.
  arXiv:0901.4085 [astro-ph.HE].

\bibitem{Albuquerque:2010rq}
  Albuquerque IFM, Chou A.
  arXiv:1001.0972 [astro-ph.HE].

\bibitem{Payez:2008pm}
  Payez A, Cudell JR, Hutsemekers D.
  {\it AIP Conf.\ Proc.\ } 1038:211 (2008)
  [arXiv:0805.3946 [astro-ph]]

\bibitem{Zioutas:2008ie}
  Zioutas K, Tsagri M, Semertzidis Y, Papaevangelou T.
  arXiv:0808.1545 [astro-ph]\\
  Zioutas K, Tsagri M, Papaevangelou T, Dafni T.
  {\it New J.\ Phys.\ } 11:105020 (2009)
  [arXiv:0903.1807 [astro-ph.SR]]

\bibitem{Pospelov:2007mp}
Pospelov M, Ritz A, Voloshin MB.
{\it Phys.\ Lett.\  B} 662:53 (2008)
[arXiv:0711.4866 [hep-ph]]

\bibitem{Ibarra:2008kn}
Ibarra A, Ringwald A, Weniger C.
{\it JCAP} 0901:003 (2009) 003
[arXiv:0809.3196 [hep-ph]]

\bibitem{Shirai:2009kh}
Shirai S, Takahashi F, Yanagida TT.
arXiv:0902.4770 [hep-ph]\\
Ibarra A, Ringwald A, Tran D, Weniger C.
{\it JCAP} 0908:017 (2009)
[arXiv:0903.3625 [hep-ph]]



\bibitem{ArkaniHamed:2008qn}
Arkani-Hamed N, Finkbeiner DP, Slatyer TR, Weiner N.
{\it Phys.\ Rev.\  D} 79:015014 (2009)
[arXiv:0810.0713 [hep-ph]]

\bibitem{Adriani:2008zr}
Adriani O, et al.   [PAMELA Collaboration].
{\it Nature} 458:607 (2009)

\bibitem{Bernabei:2008yi}
Bernabei R., et al.   [DAMA Collaboration].
{\it Eur.\ Phys.\ J.\  C} 56:333 (2008)
[arXiv:0804.2741 [astro-ph]]

\bibitem{ArkaniHamed:2008qp}
Arkani-Hamed N, Weiner N.
{\it JHEP} 0812:104 (2008)
[arXiv:0810.0714 [hep-ph]]\\
Essig R, Schuster P, Toro N.
{\it Phys.\ Rev.\  D} 80:015003 (2009)
[arXiv:0903.3941 [hep-ph]]\\
Bjorken JD, Essig R, Schuster P, Toro N.
{\it Phys.\ Rev.\  D} 80:075018 (2009)
[arXiv:0906.0580 [hep-ph]]

\bibitem{Baumgart:2009tn}
Baumgart M, et al.
{\it JHEP} 0904:014 (2009)
[arXiv:0901.0283 [hep-ph]]

\bibitem{Chun:2008by}
Chun EJ, Park JC.
{\it JCAP} 0902:026 (2009)
[arXiv:0812.0308 [hep-ph]]\\
Cheung C, Ruderman JT, Wang LT, Yavin I.
{\it Phys.\ Rev.\  D} 80:035008 (2009)
[arXiv:0902.3246 [hep-ph]]\\
Morrissey DE, Poland D, Zurek KM.
{\it JHEP} 0907:050 (2009)
[arXiv:0904.2567 [hep-ph]]\\
Cui Y, Morrissey DE, Poland D, Randall L.
{\it JHEP} 0905:076 (2009)
[arXiv:0901.0557 [hep-ph]]

\bibitem{Suematsu:2006wh}
Suematsu D.
{\it JHEP} 0611:029 (2006)
[arXiv:hep-ph/0606125]\\
  Katz A, Sundrum R.
  {\it JHEP} 0906:003 (2009)
  [arXiv:0902.3271 [hep-ph]]\\
Batell B, Pospelov M, Ritz A.
{\it Phys.\ Rev.\  D} 79:115008 (2009)
[arXiv:0903.0363 [hep-ph]]\\
Batell B, Pospelov M, Ritz A.
  {\it Phys.\ Rev.\  D} 80:095024 (2009)
  [arXiv:0906.5614 [hep-ph]]\\
Dedes A, Giomataris I, Suxho K, Vergados JD.
  {\it Nucl.\ Phys.\  B} 826:148 (2010)
  [arXiv:0907.0758 [hep-ph]]\\
Ruderman JT, Volansky T.
arXiv:0908.1570 [hep-ph]

\bibitem{Komatsu:2010fb}
  Komatsu E, et al.
  arXiv:1001.4538 [Unknown].

\bibitem{Jaeckel:2009wm}
Jaeckel J, Redondo J, Ringwald A.
{\it Europhys.\ Lett.\ } 87:10010 (2009)
[arXiv:0903.5300 [hep-ph]]

\bibitem{Okun:1982xi}
  Okun LB,
  {\it Sov.\ Phys.\ JETP} 56:502 (1982)
  [{\it Zh.\ Eksp.\ Teor.\ Fiz.\ } 83:892 (1982)]

\bibitem{Anselm:1986gz}
  Anselm AA.
  {\it Yad.\ Fiz.\ } 42:1480 (1985);
  Anselm AA.
  {\it Phys.\ Rev.\  D} 37:2001 (1988)

\bibitem{VanBibber:1987rq}
  Van Bibber K, et al.
  {\it Phys.\ Rev.\ Lett.\ } 59:759 (1987)

\bibitem{Raffelt:1987im}
  Raffelt G, Stodolsky L.
  {\it Phys.\ Rev.\  D} 37:1237 (1988)

\bibitem{Hoogeveen:1990vq}
  Hoogeveen F, Ziegenhagen T.
  {\it Nucl.\ Phys.\  B} 358:3 (1991)

\bibitem{Sikivie:2007qm}
  Sikivie P, Tanner DB, van Bibber K.
  {\it Phys.\ Rev.\ Lett.\ } 98:172002 (2007)
  [hep-ph/0701198].

\bibitem{Ahlers:2007rd}
  Ahlers M, et al.
  {\it Phys.\ Rev.\  D} 76:115005 (2007)
  [arXiv:0706.2836 [hep-ph]]

\bibitem{Ruoso:1992nx}
Ruoso G., et al.  [BFRT Collaboration].
{\it Z.\ Phys.\  C} 56:505 (1992)\\
Robilliard C, et al.  [BMV Collaboration].
{\it Phys.\ Rev.\ Lett.\ } 99:190403 (2007)
[arXiv:0707.1296 [hep-ex]]\\
Chou AS, et al.   [GammeV (T-969) Collaboration].
{\it Phys.\ Rev.\ Lett.\ } 100:080402 (2008)
[arXiv:0710.3783 [hep-ex]]\\
Afanasev A, et al.  [LIPSS Collaboration].
{\it Phys.\ Rev.\ Lett.\ } 101:120401 (2008)
[arXiv:0806.2631 [hep-ex]]\\
Fouche M, et al.  [BMV Collaboration].
{\it Phys.\ Rev.\  D} 78:032013 (2008)
[arXiv:0808.2800 [hep-ex]]\\
  Afanasev A, et al.  [LIPSS Collaboration].
  {\it Phys.\ Lett.\  B} 679:317 (2009)
  [arXiv:0810.4189 [hep-ex]]\\
  Ehret K, et al.   [ALPS collaboration].
  {\it Nucl.\ Instrum.\ Meth.\  A} 612:83 (2009)
  [arXiv:0905.4159 [physics.ins-det]]

\bibitem{Cameron:1993mr}
Cameron R, et al.  [BFRT Collaboration].
{\it Phys.\ Rev.\  D} 47:3707 (1993);

\bibitem{ALPS}
  Lindner A, Redondo J, for the [ALPS Collaboration]. Private communication
  and publication in preparation


\bibitem{cantatore}
  Cantatore G. Talk at the 5th Patras Workshop on Axions, WIMPs and WISPs, http://axion-wimp.desy.de\\
  Mueller G, Sikivie P, Tanner DB, van Bibber K.
  {\it Phys.\ Rev.\  D} 80:072004 (2009)
  [arXiv:0907.5387 [hep-ph]]

\bibitem{Jaeckel:2007gk}
  Jaeckel J, Ringwald A.
  {\it Phys.\ Lett.\  B} 653:167 (2007)
  [arXiv:0706.0693 [hep-ph]]

\bibitem{Rabadan:2005dm}
  Buchmuller W, Hoogeveen F.
  {\it Phys.\ Lett.\  B} 237:278 (1990)\\
%
  Rabadan R, Ringwald A, Sigurdson K.
  {\it Phys.\ Rev.\ Lett.\ } 96:110407 (2006)
  [arXiv:hep-ph/0511103]\\
  Dias AG, Lugones G.
  {\it Phys.\ Lett.\  B} 673:101 (2009)
  [arXiv:0902.0749 [hep-ph]]

\bibitem{Hoogeveen:1992uk}
  Hoogeveen F.
  {\it Phys.\ Lett.\  B} 288:195 (1992)\\
  Jaeckel J, Ringwald A.
  {\it Phys.\ Lett.\  B} 659:509 (2008)
  [arXiv:0707.2063 [hep-ph]]\\
  Caspers F, Jaeckel J., Ringwald A.
  {\it JINST} 4:P11013 (2009)
  [arXiv:0908.0759 [hep-ex]].

\bibitem{Yale}
  Slocum P. Talk at the 5th Patras Workshop on Axions, WIMPs and WISPs, http://axion-wimp.desy.de\\
  Williams P. Talk at the 5th Patras Workshop on Axions, WIMPs and WISPs, http://axion-wimp.desy.de\\
  Tobar M. Talk at the 5th Patras Workshop on Axions, WIMPs and WISPs, http://axion-wimp.desy.de

\bibitem{ADMXcham}
  Rybka G. Talk at the AXIONS 2010 Workshop, http://www.phys.ufl.edu/research/Axions2010/

\bibitem{Jaeckel:2008sz}
  Jaeckel J, Redondo J.
  {\it Europhys.\ Lett.\ } 84:31002 (2008)
  [arXiv:0806.1115 [hep-ph]]

\bibitem{Ahlers:2007st}
  Ahlers M, et al.
  {\it Phys.\ Rev.\  D} 77:015018 (2008)
  [arXiv:0710.1555 [hep-ph]]

\bibitem{Gies:2007su}
  Gies H, Mota DF, Shaw DJ.
  {\it Phys.\ Rev.\  D} 77:025016 (2008)
  [arXiv:0710.1556 [hep-ph]]

\bibitem{Chou:2008gr}
  Chou AS, et al.   [GammeV Collaboration].
  {\it Phys.\ Rev.\ Lett.\ } 102:030402 (2009)
  [arXiv:0806.2438 [hep-ex]]

\bibitem{gammev}
  Wester W. Poster presented at the AXIONS 2010 Workshop, http://www.phys.ufl.edu/research/Axions2010/

\bibitem{Sikivie:1983ip}
  Sikivie P.
  {\it Phys.\ Rev.\ Lett.\ } 51:1415 (1983)
  [{\it Erratum-ibid.\ } 52:695 (1984)]

\bibitem{vanBibber:1988ge}
  van Bibber K, McIntyre PM, Morris DE, Raffelt GG.
  {\it Phys.\ Rev.\  D} 39:2089 (1989)\\
  Paschos EA, Zioutas K.
  {\it Phys.\ Lett.\  B} 323:367 (1994)

\bibitem{karuza}
  Karuza M. Talk at the 5th Patras Workshop on Axions, WIMPs and WISPs, http://axion-wimp.desy.de

\bibitem{Gninenko:2008pz}
  Gninenko SN, Redondo J.
  {\it Phys.\ Lett.\  B} 664:180 (2008)
  [arXiv:0804.3736 [hep-ex]]

\bibitem{ferrer}
  Ferrer-Ribas E. Talk at the 5th Patras Workshop on Axions, WIMPs and WISPs, http://axion-wimp.desy.de

\bibitem{DedHelio}
  Ohta R. Talk at the 5th Patras Workshop on Axions, WIMPs and WISPs, http://axion-wimp.desy.de

\bibitem{SHIPS}
Redondo J. Talk at PASCOS 2009, http://pascos2009.desy.de/

\bibitem{Davoudiasl:2005nh}
  Davoudiasl H, Huber P.
  {\it Phys.\ Rev.\ Lett.\ } 97:141302 (2006)
  [arXiv:hep-ph/0509293];
  {\it JCAP} 0808:026 (2008)
  [arXiv:0804.3543 [astro-ph]]

\bibitem{Fairbairn:2007vj}
  Fairbairn M, et al.
  {\it Eur.\ Phys.\ J.\  C} 52:899 (2007)
  [arXiv:0706.0108 [hep-ph]]\\
  Fairbairn M, Rashba T, Troitsky SV.
  {\it Phys.\ Rev.\ Lett.\ } 98:201801 (2007)
  [arXiv:astro-ph/0610844]


\bibitem{Duffy:2006aa}
  Duffy LD, et al.
  {\it Phys.\ Rev.\  D} 74:012006 (2006)
  [arXiv:astro-ph/0603108]\\
  Asztalos SJ, et al.   [The ADMX Collaboration].
  arXiv:0910.5914 [astro-ph.CO]

\bibitem{Tada:1999tu}
  Tada M, et al.
  {\it Nucl.\ Phys.\ Proc.\ Suppl.\ } 72:164 (1999)

\bibitem{Melissinos:2008vn}
  Melissinos AC.
  {\it Phys.\ Rev.\ Lett.\ } 102:202001 (2009)
  [arXiv:0807.1092 [hep-ph]]

\bibitem{Thomas}
  Thomas S. Talk at the AXIONS 2010 workshop, http://www.phys.ufl.edu/research/Axions2010/

\bibitem{Gies:2006ca}
  Gies H, Jaeckel J, Ringwald A.
  {\it Phys.\ Rev.\ Lett.\ } 97:140402 (2006)
  [arXiv:hep-ph/0607118]

\bibitem{Maiani:1986md}
  Maiani L, Petronzio R, Zavattini E.
  {\it Phys.\ Lett.\  B} 175:359 (1986) 359

\bibitem{Brandi:2000ty}
  Brandi F, et al.
  {\it Nucl.\ Instrum.\ Meth.\  A} 461:329 (2001)
  [arXiv:hep-ex/0006015]

\bibitem{Gies:2009wx}
  Gies H, Jaeckel J.
  {\it JHEP} 0908:063 (2009) 063
  [arXiv:0904.0609 [hep-ph]]

\bibitem{Ahlers:2008jt}
  Ahlers M, Jaeckel J, Ringwald A.
  {\it Phys.\ Rev.\  D} 79:075017 (2009)
  [arXiv:0812.3150 [hep-ph]]

\bibitem{Zavattini:2005tm}
  Zavattini E, et al.   [PVLAS Collaboration].
  {\it Phys.\ Rev.\ Lett.\ } 96:110406 (2006)
  [{\it Erratum-ibid.\ } 99:129901 (2007)]
  [arXiv:hep-ex/0507107]\\
  Zavattini E, et al.   [PVLAS Collaboration].
  {\it Phys.\ Rev.\  D} 77:032006 (2008)
  [arXiv:0706.3419 [hep-ex]]\\
  Chen SJ, Mei HH, Ni WT.
  {\it Mod.\ Phys.\ Lett.\  A} 22:2815 (2007)
  [arXiv:hep-ex/0611050]\\
Battesti R, et al.
{\it Eur. Phys. J. D} 46:323 (2008)
[arXiv:0710.1703]\\
Pugnat P, et al.  [OSQAR Collaboration].
Optical search for QED vacuum magnetic birefringence, axions and photon regeneration (OSQAR),
CERN-SPSC-2006-035, CERN-SPSC-P-331\\
  Cantatore G, et al.
  [arXiv:0809.4208 [hep-ex]]

\bibitem{Dobrich:2009kd}
  Boer D, Van Holten JW.
  arXiv:hep-ph/0204207\\
  Zavattini G, Calloni E.
  {\it Eur.\ Phys.\ J.\  C} 62:459 (2009)
  [arXiv:0812.0345 [physics.ins-det]]\\
  Dobrich B, Gies H.
  {\it Europhys.\ Lett.\ } 87:21002 (2009)
  [arXiv:0904.0216 [hep-ph]]\\
  Dobrich B, Gies H.
  arXiv:0910.5692 [hep-ph]

\bibitem{Guendelman:2008jm}
  Guendelman EI.
  {\it Phys.\ Lett.\  B} 662:445 (2008)
  [arXiv:0802.0311 [hep-th]]\\
Redondo K. Private communication

\bibitem{Schwinger:1951nm}
Schwinger J.
{\it Phys.\ Rev.\ } 82:664 (1951)

\bibitem{Gies:2006hv}
  Gies H, Jaeckel J, Ringwald A.
  {\it Europhys.\ Lett.\ } 76:794 (2006)
  [arXiv:hep-ph/0608238]

\bibitem{Badertscher:2006fm}
  Badertscher A, et al.
  {\it Phys.\ Rev.\  D} 75:032004 (2007)
  [arXiv:hep-ex/0609059]\\
  Ahlers M, et al.
  {\it Phys.\ Rev.\  D} 77:095001 (2008)
  [arXiv:0711.4991 [hep-ph]]


\bibitem{Jaeckel:2009dh}
  Jaeckel J.
  {\it Phys.\ Rev.\ Lett.\ } 103:080402 (2009)
  [arXiv:0904.1547 [hep-ph]]

\bibitem{Adelberger:2009zz}
  Adelberger EG, et al.
  {\it Prog.\ Part.\ Nucl.\ Phys.\ } 62:102 (2009)

\bibitem{Moody:1984ba}
  Moody JE, Wilczek F.
  {\it Phys.\ Rev.\  D} 30:130 (1984)

\bibitem{Ferrer:1998ue}
  Ferrer F, Grifols JA.
  {\it Phys.\ Rev.\  D} 58:096006 (1998)
  [arXiv:hep-ph/9805477]

\bibitem{Adelberger:2006dh}
  Adelberger EG, et al.
  {\it Phys.\ Rev.\ Lett.\ } 98:131104 (2007)
  [arXiv:hep-ph/0611223]

\bibitem{Dupays:2006dp}
  Dupays A, Masso E, Redondo J, Rizzo C.
  {\it Phys.\ Rev.\ Lett.\ } 98:131802 (2007)
  [arXiv:hep-ph/0610286]

\bibitem{Su:1994gu}
  Su Y, et al.
  {\it Phys.\ Rev.\  D} 50:3614 (1994)

\bibitem{Spero:1980zz}
  Spero R, et al.
  {\it Phys.\ Rev.\ Lett.\ } 44:1645 (1980)\\
%
  Hoskins JK, Newman RD, Spero R, Schultz J.
  {\it Phys.\ Rev.\  D} 32:3084 (1985)

\bibitem{Kapner:2006si}
  Kapner DJ, et al.
  {\it Phys.\ Rev.\ Lett.\ } 98:021101 (2007)
  [arXiv:hep-ph/0611184]

\bibitem{Chiaverini:2002cb}
  Chiaverini J, et al.
  {\it Phys.\ Rev.\ Lett.\ } 90:151101 (2003)
  [arXiv:hep-ph/0209325]\\
%
  Smullin SJ, et al.
  {\it Phys.\ Rev.\  D} 72:122001 (2005)
  [{\it Erratum-ibid.\  D} 72:129901 (2005)]
  [arXiv:hep-ph/0508204]

\bibitem{Lamoreaux:1996wh}
  Lamoreaux SK.
  {\it Phys.\ Rev.\ Lett.\ } 78:5 (1997)
  [{\it Erratum-ibid.\ } 81:5475 (1998)]

\bibitem{Dickey:1994}
  Dickey JO, et al.
  {\it Science} 265:482 (1994)\\
%
  Williams JG, Newhall XX, Dickey JO.
  {\it Phys.\ Rev.\  D} 53:6730 (1996)

\bibitem{Onofrio:2006mq}
  Onofrio R.
  {\it New J.\ Phys.\ } 8:237 (2006)
  [arXiv:hep-ph/0612234]

\bibitem{Vasilakis:2008yn}
  Vasilakis G, Brown JM, Kornack TW, Romalis MV.
  {\it Phys.\ Rev.\ Lett.\ } 103:241801 (2009)
  [arXiv:0809.4700 [physics.atom-ph]]

\bibitem{Popov:1999}
  Popov V.
  {\it Turk.\ J.\ Phys.\ } 23:943 (1999)

\bibitem{Williams:1971ms}
  Williams ER, Faller JE, Hill HA.
  {\it Phys.\ Rev.\ Lett.\ } 26:721 (1971)


\bibitem{Goldhaber:1971mr}
  Goldhaber AS, Nieto MM.
  {\it Rev.\ Mod.\ Phys.\ } 43:277 (1971)

\bibitem{Bartlett:1988yy}
  Bartlett DF, Loegl S.
  {\it Phys.\ Rev.\ Lett.\ } 61:2285 (1988)

\bibitem{Goldhaber:1968mt}
  Goldhaber AS, Nieto MM.
  {\it Phys.\ Rev.\ Lett.\ } 21:567 (1968)

\bibitem{Kostelecky:2007qf}
  Kostelecky VA.
  arXiv:0802.0581 [gr-qc]\\
  Kostelecky VA, Mewes M.
  {\it Phys.\ Rev.\  D} 66:056005 (2002)
  [arXiv:hep-ph/0205211]

\bibitem{Chelouche:2008ta}
  Chelouche D, Rabadan R, Pavlov S, Castejon F.
  {\it Astrophys.\ J.\ Suppl.\ } 180:1 (2009)
  [arXiv:0806.0411 [astro-ph]]\\
%
  Bassan N, Mirizzi A, Roncadelli M.
  arXiv:1001.5267 [astro-ph.HE]

\end{thebibliography}
\end{document}